\documentclass[a4paper,11pt]{article}
\pdfoutput=1 
\usepackage{jheppub} 
\usepackage[T1]{fontenc} 
\usepackage[utf8]{inputenc} 
\usepackage{amsmath}
\usepackage{amssymb}
\usepackage{epsfig}
\usepackage{graphicx}
\usepackage{float}
\usepackage{booktabs}

\newcommand{\psl}{ p \hspace{-2.0truemm}/ }
\newcommand{\epsl}{ \epsilon \hspace{-2.0truemm}/ }

\title{\boldmath The $\rho(770,1450)\to \omega\pi$ contributions for three-body decays $B\to\bar{D}^{(*)} \omega\pi$} 

\author[a,c]{Yu-Shan Ren}
\author[b]{Ai-Jun Ma}
\author[d,a,1]{and Wen-Fei Wang,\note{Corresponding author.}}

\affiliation[a]{Institute of Theoretical Physics and State Key Laboratory of Quantum Optics and Quantum Optics Devices, 
                     Shanxi University, 
                     \\Taiyuan, Shanxi 030006, China}
\affiliation[b]{School of Mathematics and Physics,  Nanjing Institute of Technology,\\Nanjing, Jiangsu 211167, China}
\affiliation[c]{School of Physics, University of Electronic Science and Technology of China, \\Chengdu 610054, China}
\affiliation[d]{Departament de F\'{\i}sica Qu\`antica i Astrof\'{\i}sica and Institut de Ci\`encies del Cosmos (ICCUB), 
                         Facultat de F\'{\i}sica, Universitat de Barcelona, \\Mart\'i i Franqu\`es 1, 08028, Barcelona, Spain }
		         
\emailAdd{wfwang@ub.edu} 

\abstract{
The decays $B\to\bar{D}^{(*)} \omega\pi$ are very important for the investigation of $\rho$ excitations and the 
test of factorization hypothesis for $B$ meson decays. The $B^{+}\to \bar{D}^{(*)0}\omega\pi^+$ and 
$B^{0}\to D^{(*)-}\omega\pi^+$  have been measured by different collaborations but without any predictions 
for their observables on theoretical side. In this work, we study the contributions of $\rho(770,1450)\to \omega\pi$ 
for the cascade decays $B^{+}\to \bar{D}^{(*)0} \rho^+ \to \bar{D}^{(*)0}\omega\pi^+$, 
$B^{0}\to D^{(*)-} \rho^+ \to D^{(*)-}\omega\pi^+$ and $B_s^{0}\to D_s^{(*)-} \rho^+ \to D^{(*)-}\omega\pi^+$. 
We introduce $\rho(770,1450)\to \omega\pi$ subprocesses into the distribution amplitudes for $\omega\pi$ 
system via the vector form factor $F_{\omega\pi}(s)$ and then predict the branching fractions for the first time 
for concerned quasi-two-body decays with $\rho(770,1450)\to \omega\pi$, as well as the corresponding 
longitudinal polarization fractions $\Gamma_L/\Gamma$ for the cases with the vector $\bar{D}^{*0}$ or 
$D_{(s)}^{*-}$ in their final states. The branching fractions of these quasi-two-body decays are predicted at 
the order of $10^{-3}$, which can be detected at the LHCb and Belle-II experiments. 
The predictions for the decays ${B}^0 \to{D}^{*-} \rho(770)^+\to {D}^{*-} \omega\pi^+$ and 
${B}^0 \to {D}^{*-} \rho(1450)^+\to {D}^{*-} \omega\pi^+$ agree well with the measurements from Belle  
Collaboration. In order to avoid the pollution from annihilation Feynman diagrams, we recommend to 
take the $B_s^0 \to D_s^{*-}\rho(770,1450)^+$ decays, which have only emission diagrams 
at quark level, to test the factorization hypothesis for $B$ decays.
}

\begin{document} 
\maketitle
\flushbottom

\section{Introduction}

Three-body hadronic $B$ meson decay processes always provide us a rich field to investigate various aspects of the 
strong and weak interactions. We may rely on them to study dynamical models for the strong interaction, to analyse hadron 
spectroscopy and explore the properties and substructures of resonant states, to determine the fundamental parameters 
for quark mixing and to understand the essence of $CP$ asymmetries. In recent years, experimental efforts on these  
decay processes by employing Dalitz plot technique~\cite{prd94.1046} have revealed valuable insights into the 
involved strong and weak dynamics. But on the theoretical side, it is complicated to describe the strong dynamics 
in these decays because of the rescattering processes~\cite{1512.09284,prd89.094013,epjc78-897,prd71.074016}, 
hadron-hadron interactions and three-body effects~\cite{npps199-341,prd84.094001} in the final states. The resonance 
contributions in relevant decay channels, which are associated with the scalar, vector and tensor intermediate states, 
could be isolated from the total decay amplitudes and can be studied within the quasi-two-body framework~\cite{plb763-29,
1605.03889,prd96.113003}.

Three-body decays $B\to\bar{D}^{(*)} \omega\pi$, with one open charm meson in the final state of each channel, 
are relatively simple from the theoretical point of view. The heavy $b$-quark weak decay in these processes 
receive contributions only from tree-level $W$ exchange operators $O_1$ and $O_2$, which can be described well 
by the effective Hamiltonian $\mathcal{H}_{\rm eff}$~\cite{rmp68.1125} within the factorization 
method~\cite{zpc34-103}. Among these decays, $B^+ \to \bar{D}^{(*)0} \omega\pi^+$ and 
${B}^0 \to {D}^{(*)-} \omega\pi^+$ were measured by CLEO Collaboration for the first time twenty years 
ago~\cite{prd64.092001}. The decay ${B}^0 \to {D}^{*-} \omega\pi^+$ was studied later by {\it BABAR} 
and Belle Collaborations with the updated total branching fractions 
$(2.88\pm0.21({\rm stat.})\pm0.31({\rm syst.}))\times10^{-3}$~\cite{prd74.012001} and                
$(2.31\pm0.11({\rm stat.})\pm0.14({\rm syst.}))\times10^{-3}$~\cite{prd92.012013}, respectively.  
The $\omega\pi$ pair in the final states of $B\to\bar{D}^{(*)} \omega\pi$ decays is related to the resonance 
$\rho(1450)$, the excitation of $\rho(770)$~\cite{PDG22}. In $B$ meson decays, $\rho(1450)$ was actually 
observed for the first time in $B\to\bar{D}^{(*)} \omega\pi$ decays by CLEO in~\cite{prd64.092001}. 
In Ref.~\cite{prd92.012013}, the surprising large contribution for $\omega\pi$ from $\rho(770)$ in 
${B}^0 \to {D}^{*-} \omega\pi^+$ decay was measured to be 
\begin{eqnarray}
    \mathcal{B}({B}^0 \to {D}^{*-} \rho(770)^+\to  {D}^{*-} \omega\pi^+) 
                                  =(1.48^{+0.37}_{-0.63})\times10^{-3}   
\end{eqnarray}
as the branching fraction ($\mathcal{B}$), which is comparable to the corresponding data~\cite{prd92.012013,PDG22}
\begin{eqnarray}
     \mathcal{B}( {B}^0 \to {D}^{*-} \rho(1450)^+\to {D}^{*-} \omega\pi^+) 
                                   =(1.07^{+0.40}_{-0.34})\times10^{-3}   
\end{eqnarray}
for the intermediate state $\rho(1450)$.          

The natural decay mode of $\rho(770)\to \omega\pi$ is blocked as a result of the resonance pole mass which is below 
the threshold of the $\omega\pi$ pair. But the virtual contribution~\cite{plb25-294,Dalitz62,prd94.072001,plb791-342} 
from the Breit-Wigner (BW)~\cite{BW-model} tail for resonance $\rho(770)$ was found playing a vital role in 
the production of $\omega\pi$ for the processes of $e^+e^- \to \omega\pi^0$~\cite{plb174-453,plb486-29,
plb562-173,plb669-223,JETPL94-734,prd88.054013,prd94.112001,prd96.092009,plb813-136059,2309.00280} 
and $\tau \to \omega \pi \nu_\tau$~\cite{plb185-223,prd61.072003,rmp78.1043,prl103.041802}.
For the resonance $\rho(1450)$, its most precise determination of the mass and width comes actually from 
$e^+e^-$ annihilation and the related process of $\tau$ decay~\cite{zpc62-455}. The mass of $\rho(1450)$ 
is consistent with that for the $2S$ excitation of  $\rho(770)$~\cite{prd55.4157}, 
but it has been suggested as a $2S$-hybrid mixture in Ref.~\cite{prd60.114011} because of its decay 
characters~\cite{npb443-233,prd52.1706,prd56.1584}. The study of $\rho(1450)$ in $B$ decays and the 
investigation of its interference with its ground state would lead to a better understanding of its 
properties~\cite{prd92.012013}.  Its contributions for the kaon pair have been 
explored in Refs.~\cite{prd101.111901,prd103.056021,prd103.016002,cpc46-053104} and in Refs.~\cite{prd85.092016,
prd93.052018,prd103.114028,prd104.116019} in three-body $B$ and $D$ meson decays, respectively, in recent years.

In this paper, we shall concentrate on the cascade decays $B^{+}\to \bar{D}^{(*)0}\rho^+ \to \bar{D}^{(*)0} \omega\pi^+$, 
$B^{0}\to D^{(*)-}\rho^+ \to D^{(*)-}\omega\pi^+$ and $B_s^{0}\to D_s^{(*)-} \rho^+ \to D^{(*)-} \omega\pi^+$, 
where $\rho^+$ in this work stands for the intermediate states $\rho(770)^+$ and $\rho(1450)^+$ decaying 
into $\omega\pi^+$.  In the very recent study performed by SND Collaboration for 
$e^+e^- \to \omega\pi^0 \to \pi^+\pi^-\pi^0\pi^0$ process in the energy range $1.05$-$2.00$ GeV, four isovector 
vector resonances covering $\rho(770)$, $\rho(1450)$, $\rho(1700)$ and $\rho(2150)$ have been employed to 
parametrize the related form factor for the $\rho\to\omega\pi$ transition~\cite{2309.00280}. But we noticed from the
Born cross section in Ref.~\cite{2309.00280} that the contributions for $\omega\pi$ from $\rho(1700)$ and the so called 
$\rho(2150)$ state are not large and not important when comparing with those from $\rho(770)$ and $\rho(1450)$. 
In addition, the excited $\rho$ states around $2$ GeV are not well understood~\cite{plb813-136059,prd105.074035}. 
In this context we will leave the contributions for $\omega\pi$ from $\rho(1700)$ and $\rho(2150)$ in the 
concerned decays to future studies.

\begin{figure}[tbp]  
\centerline{\epsfxsize=6cm \epsffile{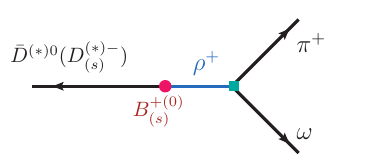}}
\caption{Schematic view of the cascade decays $B^{+}\to \bar{D}^{(*)0} \rho^+ \to \bar{D}^{(*)0} \omega\pi^+$, 
              $B^{0}\to D^{(*)-} \rho^+ \to D^{(*)-} \omega\pi^+$ and $B_s^{0}\to D_s^{(*)-} \rho^+ \to D^{(*)-} \omega\pi^+$, 
              here $\rho^+$ stands for the intermediate states $\rho(770,1450)^+$ decaying into $\omega\pi^+$ in this 
              work. }
\label{fig-1}
\vspace{-0.2cm}
\end{figure}

The schematic diagram for the cascade decays $B^{+}\to \bar{D}^{(*)0} \rho^+ \to \bar{D}^{(*)0} \omega\pi^+$, 
$B^{0}\to D^{(*)-} \rho^+ \to D^{(*)-}\omega\pi^+$ and $B_s^{0}\to D_s^{(*)-} \rho^+ \to D^{(*)-} \omega\pi^+$ is 
shown in Fig.~\ref{fig-1}. 
In the $B$ meson rest frame, the initial state will decay into the intermediate 
resonance $\rho^+$ as well as the bachelor state $\bar{D}^{(*)0}$ or $D_{(s)}^{(*)-}$, and then the resonance 
decays into its daughters $\omega$ and $\pi^+$. The state $\omega$ can be independently reconstructed from 
its two channels $\omega\to \pi^+\pi^-\pi^0$ and $\omega\to \pi^0\gamma$~\cite{plb562-173,2309.00280,
prd94.112001,prd96.092009,prd88.054013,JETPL94-734}. The decay process $B^{0} \to D^{*-}\omega\pi^+$
has only been studied in Refs.~\cite{JHEP1109-129,JHEP2002-168} with the factorization hypothesis on the 
theoretical side but without any observable predictions for its branching fraction. In this work, we shall study
these concerned cascade decays in the perturbative QCD (PQCD) approach~\cite{plb504-6,prd63.054008,
prd63.074009,ppnp51-85,prd70.054006}. The subprocesses $\rho(770,1450)\to \omega\pi$ in these decays 
can not be calculated in PQCD approach; we will introduce them into 
the distribution amplitudes for $\omega\pi$ system via the vector form factor $F_{\omega\pi}$ which has been 
measured with related processes of $e^+e^-$ annihilation and $\tau$ decay. In the first approximation 
in isobar formalism~\cite{pr135.B551,pr166.1731,prd11.3165}, one can neglect the interaction between 
$\omega\pi$ system and the corresponding bachelor state in relevant decay process, and then study the 
decays $B\to \bar{D}^{(*)} \rho(770,1450) \to \bar{D}^{(*)} \omega\pi$ within the quasi-two-body 
framework~\cite{plb763-29,1605.03889,prd96.113003}.  The quasi-two-body framework based on PQCD approach 
has been discussed in detail in~\cite{plb763-29}, which has been followed in Refs.~\cite{prd101.111901,jhep2003-162,
prd96.036014,prd95.056008,prd96.093011,npb923-54,epjc80-394,prd103.013005,prd103.096016,prd103.016002,
prd103.056021,cpc41-083105,Rui2021k,Zhang2021} for the quasi-two-body $B$ meson decays in recent years. 
For relevant works on three-body $B$ decays within the symmetries one is referred to Refs.~\cite{plb564-90,
prd72.075013,prd72.094031,prd84.056002,plb727-136,plb726-337,prd89.074043,plb728-579,prd91.014029}.
Parallel analyses within QCD factorization can be found in Refs.~\cite{JHEP2011-103,jhep2006-073,plb622-207,
plb669-102,prd79.094005,prd72.094003,prd76.094006,prd88.114014,prd102.053006,prd89.074025,prd94.094015,
npb899-247,epjc75-536,prd89.094007,prd87.076007}.

This paper is organized as follows. In Sec.~\ref{sec-2}, we give a brief introduction of the theoretical framework for 
the quasi-two-body decays $B\to \bar{D}^{(*)} \rho(770,1450) \to \bar{D}^{(*)} \omega\pi$ within PQCD approach. 
In Sec.~\ref{sec-3}, we present our numerical results of the branching fractions for 
$B^{+}\to \bar{D}^{(*)0} \rho^+ \to \bar{D}^{(*)0} \omega\pi^+$, $B^{0}\to D^{(*)-} \rho^+ \to D^{(*)-}\omega\pi^+$ and 
$B_s^{0}\to D_s^{(*)-} \rho^+ \to D^{(*)-} \omega\pi^+$ along with some necessary discussions. Summary of this work 
is given in Sec.~\ref{sec-sum}. The factorization formulae for the related decay amplitudes 
are collected in the Appendix.

\section{Framework}\label{sec:2}
\label{sec-2}  

The relevant effective weak Hamiltonian $\mathcal{H}_{\rm eff}$ for the decays $B \to \bar{D}^{(*)} \rho(770,1450)$ 
with subprocesses $\rho(770,1450)\to \omega\pi$ via the $\bar{b}\to \bar{c}$ transition is written as~\cite{rmp68.1125}
\begin{equation}  
 \mathcal H_{\rm eff}= \frac{G_F}{\sqrt2} V^*_{cb} V_{ud} \big[ C_1(\mu) O^c_1(\mu)+C_2(\mu) O^c_2(\mu) \big], \;\;
\label{eff_Ham}
\end{equation}
where $G_F=1.1663788(6)\times10^{-5}$ GeV$^{-2}$~\cite{PDG22} is the Fermi coupling constant, $V_{cb}$ and 
$V_{ud}$ are the Cabibbo-Kobayashi-Maskawa~(CKM) matrix~\cite{Cabibbo,Kobayashi} elements. The Wilson 
coefficients $C_{1,2}(\mu)$ at scale $\mu$ are always combined as $a_1=C_1+C_2/3$ and $a_2=C_2+C_1/3$. 
The detailed discussion of the evaluation for $C_{1,2}(\mu)$ in PQCD approach is found in Ref.~\cite{prd63.074009},
where one will also find the values $C_1=-0.27034$ and $C_2=1.11879$ at $m_b$ scale.
The local four-quark operators ${O}^c_{1,2}$ are the products of two $V-A$ currents, and one has  
${O^c_1}=(\bar b d)_{V-A}\,(\bar u c)_{V-A}$ and ${O^c_2}=(\bar b c)_{V-A}\,(\bar u d)_{V-A}$~\cite{rmp68.1125}.

\begin{figure}[tbp]  
\centerline{\epsfxsize=13cm \epsffile{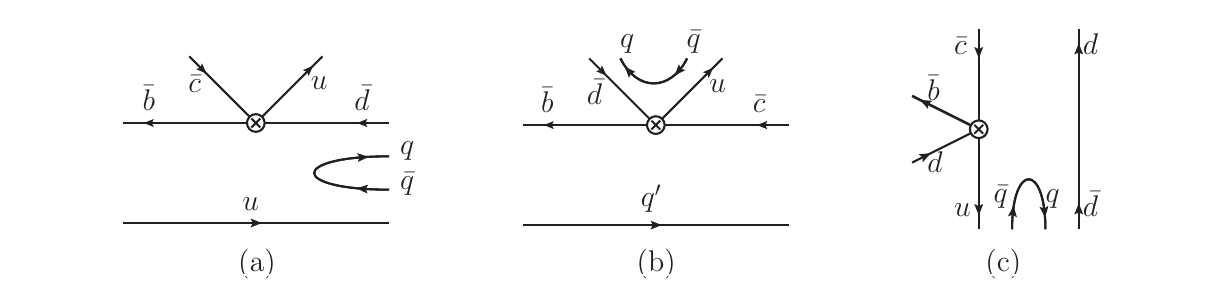}}
\caption{Typical Feynman diagrams for the quasi-two-body decays $B \to \bar{D}^{(*)} \rho\to\bar{D}^{(*)} \omega\pi$ 
              at quark level, where $q\in\{u,d\}$ and $q^\prime\in\{u, d, s\}$, the symbol $\otimes$ stands for the weak 
              interaction vertex. }
\label{fig-feyndiag}
\end{figure}

In light cone coordinates the momentum $p_B$ is equal to $\frac{m_B}{\sqrt2}(1,1,0_{\rm T})$  in the rest frame of $B$ 
meson, where the mass $m_B$ stands for initial state $B^+, B^0$ or $B^0_s$. In the 
same coordinates, the resonance $\rho(770)$, its excited state $\rho(1450)$ and the $\omega\pi$ system generated from 
resonances by strong interaction have the same momentum $p=\frac{m_B}{\sqrt 2}(\zeta, 1-r^2, 0_{\rm T})$,  with the
squared invariant mass $p^2=s$ for $\omega\pi$ system. For the bachelor state $\bar{D}^{(*)}$ in the related processes,
its momentum is defined as $p_3=\frac{m_B}{\sqrt2}(1-\zeta, r^2, 0_{\rm T})$. The longitudinal polarization vectors 
for the intermediate state and the $\bar{D}^*$ meson, respectively, are 
\begin{eqnarray}
   \epsilon^{\rho}_L&=&\frac{m_B}{\sqrt {2s}}(-\zeta, 1-r^2, 0_{\rm T}), \\
   \epsilon^{D^*}_L&=&\frac{m_B}{\sqrt2m_D}(1-\zeta, -r^2, 0_{\rm T}),
\end{eqnarray}
where the parameter $r$ will be satisfied by the relation $p^2_3=m^2_D$, with the mass $m_D$ for the bachelor state 
$\bar{D}^{(*)}$. The spectator quark comes out from initial state and goes into the intermediate states in hadronization 
shown in Fig.~\ref{fig-feyndiag}~(a) has the momenta $k_B=(\frac{m_B}{\sqrt2}x_B, 0, k_{B{\rm T}})$ and 
$k=(0, \frac{m_B}{\sqrt 2}x, k_{\rm T})$ in $B$ and $\rho$ states, respectively, and the light quark in the $\bar{D}^{(*)}$ 
got the momentum $k_3=(\frac{m_B}{\sqrt2}(1-\zeta)x_3, 0, k_{3{\rm T}})$. The $x_B$, $x$ and $x_3$, which will 
run from zero to one in the calculations, are the momentum fractions for the initial state $B$, the resonances 
$\rho(770,1450)$ and the bachelor final state $\bar{D}^{(*)}$, respectively.

In the PQCD approach,  one has factorization formula of the decay amplitude~\cite{plb561-258,prd89.074031}  
\begin{eqnarray}
     \mathcal{A}&=& \langle  (\omega\pi)_{P\text{-wave}} D^{(*)} | {\mathcal{H}_{\rm eff}} | B \rangle    \nonumber \\
                            &=&\phi_B \otimes {\mathcal H} \otimes  \phi^{P\text{-wave}} _{\omega\pi} \otimes \phi_{D^{(*)}} 
    \label{def-DA-Q2B} 
\end{eqnarray}
for the quasi-two-body decays $B \to \bar{D}^{(*)} \rho\to\bar{D}^{(*)} \omega\pi$ at leading order 
of the strong coupling $\alpha_s$ according to the Feynman diagrams of Fig.~\ref{fig-feyndiag}. Here, the hard kernel 
${\mathcal H}$ contains only one hard gluon exchange, and the symbol $\otimes$ stands for the convolutions in 
parton momenta. 

The $B$ meson light-cone matrix element in the decay amplitudes of $B\to \bar{D}^{(*)}\rho \to \bar{D}^{(*)}\omega\pi$ 
decays can be decomposed as~\cite{prd55.272,npb592-3,plb523-111}
\begin{eqnarray}
   \Phi_B=\frac{i}{\sqrt{2N_c}} (p{\hspace{-1.8truemm}/}_B+m_B)\gamma_5\phi_B (k_B),
\label{bmeson}
\end{eqnarray}
where the distribution amplitude $\phi_B$ is of the form
\begin{eqnarray}
   \phi_B(x_B,b_B) &=& N_B x_B^2(1-x_B)^2 %
                           \mathrm{exp}\left[-\frac{(x_Bm_B)^2}{2\omega_{B}^2} -\frac{1}{2} (\omega_{B}b_B)^2\right]\!,
\label{phib}
\end{eqnarray}
with two shape parameters $\omega_B = 0.40 \pm 0.04$ GeV for $B^{\pm,0}$ and 
$\omega_{B_s}=0.50 \pm 0.05$ for $B^0_s$, respectively, the $N_B$ is a normalization factor. 

The wave functions for $\bar{D}^{*0}$ and $D^{*-}_{(s)}$ have been discussed in detail in Ref.~\cite{prd78.014018}. 
Up to twist-3 accuracy, their two-particle light-cone distribution amplitudes are defined as
\begin{eqnarray}
  &&  \langle D_{(s)}(p)|q _\alpha(z)\bar c_\beta(0)|0\rangle= \frac{i}{\sqrt{2N_C}}\int_0^1dx e^{ixp\cdot z}   
                     \left[\gamma_5\left(\psl+m\right) \phi_{D_{(s)}}(x,b)\right]_{\alpha\beta}\!,    \quad          \\
  && \langle D_{(s)}^*(p)|q_\alpha(z)\bar c_\beta(0)|0\rangle = -\frac{1}{\sqrt{2N_C}}\int_0^1dx e^{ixp\cdot z}                        \big[\epsl_L(\psl+m)\phi^L_{D^*_{(s)}}(x,b)          \nonumber \\ 
  & &  \phantom{ \langle D_{(s)}^*(p)|q_\alpha(z)\bar c_\beta(0)|0\rangle }                      
                       +\epsl_T(\psl+m)\phi^T_{D^*_{(s)}}(x,b)\big]_{\alpha\beta},
\end{eqnarray}
with the normalization conditions
\begin{eqnarray}
    \int_0^1 dx\phi_D(x)        &=&\frac{f_D}{2\sqrt{2N_c}},                \\
   \int_0^1dx\phi_{D^*}^L(x) &=&\frac{f_{D^*}}{2\sqrt{2N_c}},           \\
   \int_0^1dx\phi_{D^*}^T(x) &=&\frac{f_{D^*}^T}{2\sqrt{2N_c}}.
\end{eqnarray}
The distribution amplitude for the $\bar{D}^{0}$ and $D^-_{(s)}$ mesons is~\cite{prd78.014018,prd90.094018}
\begin{eqnarray}
   \phi_{D_{(s)}}&=&\frac{1}{2\sqrt{2N_C}}f_{D_{(s)}}6x(1-x)       %
         \left[1+C_{D_{(s)}}(1-2x)\right]\exp\left[-\frac{\omega_{D_{(s)}}^2b^2}{2}\right]\!.
\end{eqnarray}    
We adopt the same structure of the distribution amplitude for both $D_{(s)}$ and $D^*_{(s)}$ in view the detailed 
discussions in Ref.~\cite{prd78.014018} for them, but we employ different Gegenbauer moments $C_D=0.6\pm0.15$ 
and $C_{D^*}=0.5\pm0.10$ for $D_{(s)}$ and $D^*_{(s)}$, respectively, 
in order to cater to the existing experimental data and also taking into account 
the different decay constants for them as they in this work and in Ref.~\cite{prd78.014018}.  

For the $P$-wave $\omega\pi$ system along with the subprocess $\rho\to \omega\pi$, the distribution amplitudes 
hold the same structure of the vector mesons and could be organized into~\cite{prd76.074018,prd101.111901,
prd102.056017}  
\begin{eqnarray} 
  \phi^{P\text{-wave}}_{\omega\pi,L}(x,s)&=&\frac{-1}{\sqrt{2N_c}}
      \big[\sqrt{s}\,{\epsilon\hspace{-1.5truemm}/}\!_L\phi^0(x,s) 
             + {\epsilon\hspace{-1.5truemm}/}\!_L {p\hspace{-1.7truemm}/} \phi^t(x,s)   
           + \sqrt s \phi^s(x,s)  \big]\!, \\
   \phi^{P\text{-wave}}_{\omega\pi,T}(x,s)&=&\frac{-1}{\sqrt{2N_c}}
      \big[\sqrt{s}\,{\epsilon\hspace{-1.5truemm}/}\!_T\phi^v(x,s) 
             + {\epsilon\hspace{-1.5truemm}/}\!_T {p\hspace{-1.7truemm}/} \phi^T(x,s)  
        +\sqrt{s}\, i\epsilon_{\mu\nu\rho\sigma}\gamma_5\gamma^\mu\epsilon_T^{*\nu} n^\rho v^\sigma \phi^a(x) \big]\!,  
       \quad      
\end{eqnarray}
with two dimensionless lightlike vectors $n=(1,0,{\bf 0}_T)$ and $v=(0,1,{\bf 0}_T)$, and $N_c$ is the number of colors for 
QCD. We adopt the convention $\epsilon^{0123}=1$ for the Levi-Civita tensor $\epsilon^{\mu\nu\alpha\beta}$.
The twist-$2$ distribution amplitude for a longitudinally polarized $\rho$ state can be parametrized as~\cite{prd76.074018}
\begin{eqnarray}
   \phi^{0}(x,s)&=&\frac{3f_{\omega\pi}(s)}{\sqrt{2N_c}} x(1-x)\big[1+a_R^{0} C^{3/2}_2(1-2x) \big].~~\label{def-DA-0}
\end{eqnarray} 
where the Gegenbauer polynomial $C^{3/2}_2(t)=3/2(5t^2-1)$. 
The twist-$2$ transversely polarized distribution amplitude $\phi^T(x,s)$ has a similar form as the longitudinally 
polarized one, we have~\cite{prd76.074018}
\begin{eqnarray}
   \phi^{T}(x,s)&=&\frac{3f_{\omega\pi}^T(s)}{\sqrt{2N_c}}x(1-x)\big[1+a^{T}_{R} C^{3/2}_2(1-2x)\big].\quad\label{def-DAT}
\end{eqnarray}
The forms of the twist-3 distribution amplitudes are~\cite{prd76.074018,prd101.111901,prd102.056017} 
\begin{eqnarray}
   \phi^{t}(x,s)&=&\frac{3f_{\omega\pi}^T(s)}{2\sqrt{2N_c}}(1-2x)^2     
                                             \big[1+a_R^t  C^{3/2}_2(1-2x)\big],\label{def-DA-t}\\
   \phi^{s}(x,s)&=&\frac{3f_{\omega\pi}^T(s)}{2\sqrt{2N_c}}(1-2x)     
                                                  \big[1+a_R^s(1-10x  +10x^2) \big],\label{def-DA-s}\\
   \phi^v(x,s)&=&\frac{3f_{\omega\pi}(s)}{8\sqrt{2N_c}}\left[1+(1-2x)^2\right]\!,\label{def-DAv}\\
   \phi^a(x,s)&=&\frac{3f_{\omega\pi}(s)}{4\sqrt{2N_c}}(1-2x). \label{def-D_phia}
\end{eqnarray}
We adopt the same Gegenbauer moments for the $P$-wave $\omega\pi$ system in this work as they were in 
Refs.~\cite{prd101.111901,plb763-29,prd103.056021} for the pion pair or kaon pair in view of the fact 
that these parameters are employed to describe the formation rather than the decay for the intermediate states. And the value 
of Gegenbauer moment $a^{T}_{R}$ in twist-$2$ transversely polarized distribution amplitude $\phi^T(x,s)$ 
is set to be the same as it for $a^{0}_{R}$ in this work. The form factor $F_{\omega\pi}^T$ for the twist-$3$ 
distribution amplitudes of $\phi^{P\text{-wave}}_{\omega\pi,L}(x,s)$
and the twist-$2$ of $\phi^{P\text{-wave}}_{\omega\pi,T}(x,s)$ are deduced from the relation 
$f_{\omega\pi}^T(s)\approx (f^T_{\rho}/f_{\rho})f_{\omega\pi}(s)$~\cite{plb763-29} with the result 
$f^T_\rho/f_\rho=0.687$ at the scale $\mu=2$ GeV~\cite{prd78.114509}. 

The factor $f_{\omega\pi}(s)$ in Eq.~(\ref{def-DA-0}) is employed as the abbreviation of the transition form factor for 
$\rho(770,1450)\to \omega\pi$ decays in the concerned processes. The related effective Lagrangian is written 
as~\cite{plb66-165,prd30.594,prd46.1195}
\begin{eqnarray}
       \mathcal{L}_{\rho\omega\pi}=g_{\rho\omega\pi}
                     \epsilon_{\mu\nu\alpha\beta}\partial^{\mu}\rho^{\nu}\partial^{\alpha}\omega^{\beta}\pi.
        \label{eq-Lagrangian}
\end{eqnarray}
With the help of this Lagrangian, we can define the form factor $F_{\omega\pi}(s)$ 
from the matrix element~\cite{epja38-331,prd92.014014,2307.10357}
\begin{eqnarray}
        & & \langle \omega(p_a,\lambda)\pi(p_b)| j_\mu(0) | 0 \rangle    %
            = i  \epsilon_{\mu\nu\alpha\beta}\varepsilon^{\nu}(p_a, \lambda)p_b^\alpha p^\beta F_{\omega\pi}(s),
  \label{eq-def-FOpi}  
\end{eqnarray}
where $j_\mu$ is the isovector part of the electromagnetic current, $\lambda$ and $\varepsilon$ is the polarization 
and polarization vector for $\omega$ meson, $p_a$ and $p_b$ are the momenta for $\omega$ and pion, respectively, 
and $p=p_a+p_b$.   We need to stress that, in order to make the expression of differential branching fraction the 
Eq.~(\ref{eqn-diff-bra}) brief and concise, we employ $f_{\omega\pi}=f^2_\rho/m_\rho F_{\omega\pi}$ to describe 
the distribution amplitudes above for the $P$-wave $\omega\pi$ system in Eqs.~(\ref{def-DA-0})-(\ref{def-D_phia}).

In the vector meson dominance model, the form factor $F_{\omega\pi}(s)$ defined by Eq.~(\ref{eq-def-FOpi}) is 
parametrized as~\cite{plb486-29,prd88.054013,prd94.112001,ppnp120-103884}
\begin{eqnarray}
	F_{\omega\pi}(s) = \frac{g_{\rho\omega\pi}}{f_\rho} \sum\limits_{\rho_i}  
	               \frac{A_i e^{i \phi_i} m_{\rho_i}^2 }{D_{\rho_i}(s)}\,,
  \label{exp-formfactor}
\end{eqnarray}   
where the summation is over the isovector resonances $\rho_i=\{\rho(770), \rho(1450), \rho(1700), ...\}$ in $\rho$ 
family, $A_i$, $\phi_i$ and $m_{\rho_i}$ are the weights, phases and masses for these resonances, respectively, 
and one has $A=1$ and $\phi=0$ for $\rho(770)$. Contributions from the excitations of $\omega$ meson can also 
be include in Eq.~(\ref{exp-formfactor}), but their weights turn out to be negligibly small~\cite{jpg36-085008}.
The parameter $f_\rho$ is the $\gamma^*\to \rho(770)$ coupling constant calculated from the decay width of 
$\rho(770)\to e^+e^-$, the $g_{\rho\omega\pi}$ is the coupling constant for $\rho(770)\to\omega\pi$ which can 
be calculated from the decay width of $\omega\to\pi^0\gamma$~\cite{PDG22} or be estimated with the relation 
$g_{\rho\omega\pi}\approx3g_{\rho\pi\pi}^2/(8\pi^2F_\pi)$~\cite{prc83.048201}, where $F_\pi=f_\pi/\sqrt2$ and 
$f_\pi$ is the decay constant for pion.  The denominator $D_{\rho_i}$ has a BW formula expression
\begin{equation}
    D_{\rho_i}(s) = m_{\rho_i}^2-s-i \sqrt{s}\,\Gamma_{\rho_i}(s).
  \label{BW-denom}
\end{equation}
To describe the shape of the resonance $\rho(770)$, the energy-dependent width is written 
as~\cite{2309.00280,prd88.054013}
\begin{eqnarray}
	\Gamma_{\rho(770)}(s) &=& \Gamma_{\rho(770)}\frac{m^2_{\rho(770)}}{s}
		\left(\frac{q_{\pi}(s)} {q_{\pi}(m^2_{\rho(770)})}\right)^{3}   
		+\frac{g_{\rho\omega\pi}^2}{12\pi} q^3_\omega(s),
\label{eq-Gm-R770}
\end{eqnarray}
where the first term of right hand side corresponds to the decay of \( \rho(770) \to \pi\pi \),  the second term is for  
\( \rho(770) \to \omega\pi \). And we have 
\begin{eqnarray}    
       && q_\pi(s) =\frac12 \sqrt{s-4m^2_\pi}\,, \label{def-qpi}   \\
     &&q_\omega(s)=\frac{1}{2\sqrt s} \sqrt{\left[s-(m_\omega+m_{\pi})^2\right]\left[s-(m_\omega-m_{\pi})^2\right]}\,.\qquad 
     \label{def-qO}           
\end{eqnarray}  
For the excited resonance $\rho(1450)$, the expression
\begin{eqnarray}    
    \Gamma_{\rho(1450)}(s) &=&\Gamma_{\rho(1450)} \Bigl[ {\mathcal B}_{\rho(1450) \to \omega\pi}  
          \Bigl(\frac {q_{\omega}(s)} {q_{\omega}(m^2_{\rho(1450)})}\Bigr)^3         \qquad    \nonumber\\
        &+&(1-{\mathcal B}_{\rho(1450) \to \omega\pi})
           \frac {m^2_{\rho(1450)}} {s}  \Bigl(\frac {q_{\pi}(s)} {q_{\pi}(m^2_{\rho(1450)})}\Bigr)^3\Bigr]
 \label{eq-Gm-R1450}
\end{eqnarray}  
for the energy-dependent width is adopted in this work as it was in Ref.~\cite{plb562-173} for the 
process $e^+ e^-\to\omega\pi^0\to\pi^0\pi^0\gamma$ by CMD-2 Collaboration, where 
${\mathcal B}_{\rho(1450) \to \omega\pi}$ is the branching ratio of the $\rho(1450) \to \omega\pi$ decay, 
$\Gamma_{\rho(770)}$ and $\Gamma_{\rho(1450)}$ are the full widths for $\rho(770)$ and $\rho(1450)$, 
respectively. 

For the differential branching fraction, one has the formula~\cite{PDG22}
\begin{eqnarray}
 \frac{d{\mathcal B}}{d s}=\tau_B\frac{s\,\vert\mathbf{p}_{\pi}\vert^3 \vert\mathbf{p}_{D}\vert^3}
                                                                 {24\pi^3m^7_B}{|{\mathcal A}|^2}
    \label{eqn-diff-bra}
\end{eqnarray}
for the quasi-two-body decays $B \to \bar{D}^{(*)} \rho\to\bar{D}^{(*)} \omega\pi$, where $\mathcal{\tau}_B$ 
is the mean lifetime for $B$ meson, $s$ is the squared invariant mass for $\omega\pi$ system. One should 
note that the phase space factor in Eq.~(\ref{eqn-diff-bra}) is different from that for the decays with subprocesses of 
$\rho \to \pi\pi$ and $\rho\to K\bar{K}$ as a result of the definition of $F_{\omega\pi}(s)$ in Eq.~(\ref{eq-def-FOpi}); 
the relations 
\begin{eqnarray}
     && \sum_{\lambda=0,\pm} \varepsilon^{\mu}{(p,\lambda)}\varepsilon^{\nu}{(p,\lambda)}=-g^{\mu\nu}
            +\frac{p^\mu p^\nu}{p^2}, \qquad \\
     && \sum_{\lambda=0,\pm}\vert\epsilon_{\mu\nu\alpha\beta}p_{3}^\mu\varepsilon^{\nu}(p_\omega, \lambda)
          p_\pi^\alpha p^\beta\vert^2  %
            = s\,\vert\mathbf{p}_{\pi}\vert^2 \vert\mathbf{p}_{D}\vert^2 (1-\cos^2{\theta})
\end{eqnarray}
are employed for the derivation of Eq.~(\ref{eqn-diff-bra}), where $\theta$ is the angle between the  three-momenta of 
$\omega$ and bachelor state $\bar{D}^{(*)}$.  In rest frame of intermediate states, the magnitude of the 
momenta are written as 
\begin{eqnarray}
   \vert\mathbf{p}_{\pi}\vert&=&\frac{\sqrt{\left[s-(m_{\pi}+m_{\omega})^2\right]\left[s-(m_{\pi}-m_{\omega})^2\right]}}{2\sqrt s}, \;\;
      \\
   \vert\mathbf{p}_{D}\vert&=&\frac{\sqrt{\left[m^2_{B}-(\sqrt s+m_{D})^2\right]\left[m^2_{B}-(\sqrt s-m_{D})^2\right]}} {2\sqrt s},\quad
 \label{eq-pKpD}
\end{eqnarray}
for pion and the bachelor meson $\bar{D}^{(*)}$, where $m_\pi$, $m_\omega$ and $m_{D}$ are the masses for 
pion, $\omega$ and the bachelor meson, respectively.  The Lorentz invariant decay amplitudes according to 
Fig.~\ref{fig-feyndiag} for the concerned decays are given in the Appendix.

\section{Results and Discussions} 
\label{sec-3}

In the numerical calculation, we employ the decay constants $f_{\rho}=0.216\pm0.003$ GeV~\cite{jhep1608-098} for 
$\rho(770)$ and $f_{\rho(1450)}=0.185^{+0.030}_{-0.035}$ GeV~\cite{plb763-29} resulting from the data~\cite{zpc62-455}
for $\rho(1450)$, the mean lives $\tau_{B^\pm}=1.638\times 10^{-12}$ s, $\tau_{B^0}=1.519\times 10^{-12}$ s and 
$\tau_{B_s^0}=1.520\times 10^{-12}$ s for the initial states $B^\pm$, $B^0$ and $B_s^0$~\cite{PDG22}, respectively.
The masses for particles in relevant decay processes, the decay constants for $B_{(s)}$, $D_{(s)}$ and 
$D^*_{(s)}$ mesons, the full widths for resonances $\rho(770)$ and $\rho(1450)$ (in units of GeV), 
and the Wolfenstein parameters for CKM matrix elements are presented in Table~\ref{tab_con}.

\begin{table}[thb] 
\begin{center}      
\caption{Masses, decay constants and full widths (in units of GeV) for relevant states as well as 
               the Wolfenstein parameters for CKM matrix elements from {\it Review of Particle Physics}~\cite{PDG22},  
               the $f_{D^*}$ and $f_{D^*_s}$ are cited from~\cite{prd96.034524}.}
\setlength{\tabcolsep}{16pt}  
\label{tab_con}
\begin{tabular}{l   r}\hline\hline
    $m_{B^{\pm}}=5.279$               &$m_{B^{0}}=5.280$          \\
    $m_{B^{0}_s}=5.367$                &$m_{D^{\pm}}=1.870$      \\
    $m_{D^{0}}=1.865$                   & $m_{D^{\pm}_s}=1.968$  \\
    $m_{D^{*\pm}}=2.010$              &$m_{D^{*0}}=2.007$          \\
    $m_{D^{*\pm}_s}=2.112$          &$m_{\pi^{\pm}}=0.140$      \\ 
    $m_{\omega}=0.783$                &$f_{B^{\pm,0}}=0.190$       \\ 
    $f_{B_s^{0}}=0.230$                  &$f_{\pi^+}=0.130$               \\   
    $f_{D^{\pm,0}}=0.2120$            &$f_{D_s^{\pm}}=0.2499$    \\     
    $f_{D^{*\pm,0}}=0.2235$          &$f_{D_s^{*\pm}}=0.2688$     \\     
 \hline                
    $m_{\rho(770)}=0.775$                          &  $m_{\rho(1450)}=1.465\pm0.025$                 \\
    $\Gamma_{\rho(770)}=0.1491$              &  $\Gamma_{\rho(1450)}=0.400\pm0.060$      \\
    $A=0.826^{+0.018}_{-0.015}$                &  $\lambda=0.22500\pm 0.00067$                   \\
 \hline\hline   
\end{tabular}
\end{center}
\end{table}   

\begin{table*}[!]
\centering
\caption{PQCD results for the quasi-two-body decays $B^{+}\to \bar{D}^{(*)0} [\rho(770)^+ \to] \pi^+\pi^0$, 
              $B^{0}\to D^{(*)-} [\rho(770)^+ \to] \pi^+\pi^0$ and $B_s^{0}\to D_s^{(*)-} [\rho(770)^+ \to] \pi^+\pi^0$,
              along with their corresponding two-body data from {\it Review of Particle Physics}~\cite{PDG22}.}
\label{tab_Rpp}   
\setlength{\tabcolsep}{12pt}  
\begin{tabular}{l c c c} \hline\hline
  \;Decay modes                                                        &\;Units            
                                                  &PQCD\;                                                                          & \;Data~\cite{PDG22} \\  
       \hline
  $B^+   \to \bar{D}^0 [\rho(770)^+ \to] \pi\pi^+$\;  &$\%$ 
                                                &\;$1.21^{+0.20}_{-0.21}$\;            & $1.34\pm0.18$\; \\ 
  $B^0   \to D^{-}     [\rho(770)^+ \to] \pi\pi^+$\;     &$10^{-3}$ 
                                                 &\;$7.63^{+1.18}_{-0.96}$\;             & $7.6\pm1.2$\; \\ 
  $B_s^0 \to D_s^{-}   [\rho(770)^+ \to] \pi\pi^+$\;  &$10^{-3}$ 
                                                 &\;$7.36^{+0.78}_{-0.82}$\;            & $6.8\pm1.4$\; \\ 
        \hline
  $B^+  \to\bar{D}^{*0}[\rho(770)^+ \to] \pi\pi^+$\;  &$10^{-3}$ 
                                                 &\;$9.03^{+1.79}_{-1.74}$\;            & $9.8\pm1.7$\; \\ 
  $B^0   \to D^{*-}    [\rho(770)^+ \to] \pi\pi^+$\;     &$10^{-3}$ 
                                                &\;$8.15^{+1.46}_{-1.45}$\;             & $6.8\pm0.9$\; \\ 
  $B_s^0 \to D_s^{*-}  [\rho(770)^+ \to] \pi\pi^+$\;  &$10^{-3}$ 
                                                &\;$7.12^{+1.09}_{-1.09}$\;             & $9.5\pm2.0$\; \\   
\hline\hline
\end{tabular}
\end{table*}

The crucial input $g_{\rho\omega\pi}$ for the form factor $F_{\omega\pi}(s)$ in Eq.~(\ref{exp-formfactor}) has been  
fitted to be $15.9\pm0.4$ GeV$^{-1}$ and $16.5\pm0.2$ GeV$^{-1}$ in~\cite{prd94.112001}, respectively, 
by SND Collaboration recently with different models for the form factor. This input can also be calculated from the 
decay width of $\omega\to \pi^0\gamma$~\cite{prd96.054033,prd56.4084}; with the relation 
$g_{\rho\omega\pi}\approx3g_{\rho\pi\pi}^2/(8\pi^2F_\pi)$~\cite{prc83.048201}, it's easy to get its value $14.8$ 
GeV$^{-1}$. In the numerical calculation of this work, we adopt $g_{\rho\omega\pi}=16.0\pm2.0$ GeV$^{-1}$ by 
taking into account the corresponding values in Refs.~\cite{prd94.112001,prd86.057301,prd61.072003,JETPL94-734,
prd55.249,prd77.113011,prd86.037302,prd88.054013} for it. The weight $A_1$ in Eq.~(\ref{exp-formfactor}) for the 
subprocess $\rho(1450)\to \omega\pi$ moves a lot in the literature, it has been measured to be $0.584\pm0.003$ 
and $0.164\pm0.003$ in~\cite{2309.00280}, $0.175\pm0.016$, $0.137\pm0.006$ and $0.251\pm0.006$ 
in~\cite{prd94.112001}, $0.26\pm0.01$ and $0.11\pm0.01$ in~\cite{prd88.054013} with different models for 
$F_{\omega\pi}(s)$ in recent years. In view of the expression for $F_{\omega\pi}(s)$ in Eq.~(\ref{exp-formfactor}), 
we have a constraint 
\begin{eqnarray}
   A_1= \frac{g_{\rho(1450)\omega\pi}f_{\rho(1450)}m_{\rho(770)}}
                               {g_{\rho(770)\omega\pi}f_{\rho(770)}m_{\rho(1450)}}
   \label{eqn-A1}
\end{eqnarray}
for its value. With the relation
\begin{eqnarray}
   f_{\rho(1450)}g_{\rho(1450)\omega\pi}
       =\sqrt{12\pi f^2_{\rho(1450)} \mathcal{B}(\rho(1450)\to\omega\pi)\Gamma_{\rho(1450)}/p_c^3}\;,
\end{eqnarray}
where $p_c=q_\omega(m^2_{\rho(1450)})$, and the measured result
$f^2_{\rho(1450)} \mathcal{B}(\rho(1450)\to\omega\pi)=0.011\pm0.003$ GeV$^2$~\cite{prd64.092001}, one has 
$A_1=0.171\pm0.036$, where the error comes from the uncertainties of mass and full width for $\rho(1450)$ in
Table \ref{tab_con}, the coupling $16.0\pm2.0$ GeV$^{-1}$ and the measured result $0.011\pm0.003$ GeV$^2$ 
in~\cite{prd64.092001}. The value for $A_1$ from Eq.~(\ref{eqn-A1}) is close to the results $0.164\pm0.003$ 
in~\cite{2309.00280} and $0.175\pm0.016$ in~\cite{prd94.112001}.

\begin{table}[thb] 
\begin{center}  
   \caption{PQCD predictions of the branching fractions (in units of $10^{-3}$) for the quasi-two-body decays 
                 $B_{(s)} \to \bar{D}_{(s)} \rho^+ \to \bar{D}_{(s)} \omega\pi^+$, where $\rho^+$ means 
                 the resonance $\rho(770)^+$ or $\rho(1450)^+$.} 
\setlength{\tabcolsep}{15pt}  
\begin{tabular}{l c} \hline\hline
  \;Decay modes                                                            
                                                 &$\mathcal{B}$ (in $10^{-3}$)\;                                                        \\  
             \hline
  $B^+   \to \bar{D}^0 [\rho(770)^+ \to] \omega\pi^+$  
                                                &\;$1.42^{+0.16+0.15+0.11+0.10}_{-0.16-0.13-0.09-0.10}$ \\ 
   $B^+   \to \bar{D}^0 [\rho(1450)^+ \to] \omega\pi^+$   
                                                &\;$0.96^{+0.11+0.09+0.08+0.40}_{-0.11-0.09-0.08-0.40}$ \\                  
              \hline                                                                            
  $B^0   \to D^{-}     [\rho(770)^+ \to] \omega\pi^+$     
                                                &\;$0.80^{+0.06+0.12+0.06+0.07}_{-0.06-0.09-0.02-0.07}$ \\ 
  $B^0   \to D^{-}     [\rho(1450)^+ \to] \omega\pi^+$     
                                                &\;$0.52^{+0.03+0.06+0.03+0.22}_{-0.03-0.06-0.03-0.22}$ \\          
             \hline                                                                           
  $B_s^0 \to D_s^{-}   [\rho(770)^+ \to]\omega\pi^+$  
                                                &\;$0.88^{+0.05+0.07+0.00+0.06}_{-0.05-0.07-0.01-0.06}$ \\ 
  $B_s^0 \to D_s^{-}   [\rho(1450)^+ \to]\omega\pi^+$ 
                                                &\;$0.59^{+0.03+0.05+0.00+0.25}_{-0.03-0.04-0.00-0.25}$ \\                                                                                              
\hline\hline
\end{tabular}\label{ResPV}  
\end{center}
\end{table}

When the subprocess $\rho(770)^+ \to \omega\pi^+$ shrink into meson $\rho(770)^+$, the six quasi-two-body 
decays of $B \to \bar{D}^{(*)} \rho(770)^+\to\bar{D}^{(*)} \omega\pi^+$ will turned into six two-body decay channels 
$B \to \bar{D}^{(*)} \rho(770)^+$. These six two-body decays with $\rho(770)^+$ have been measured, one finds their 
branching fractions in Table~\ref{tab_Rpp}. 
In view of $\mathcal{B}({\rho(770)^+\to\pi^+\pi^0})\approx 100\%$~\cite{PDG22}, the PQCD results in Table~\ref{tab_Rpp}
for the decays with subprocess $\rho(770)^+\to\pi^+\pi^0$ could be seen as a way to test the framework and inputs of 
this work. Obviously, these PQCD results in Table~\ref{tab_Rpp} agree with the data quite well.

\begin{table*}[!]
\centering
\caption{Same as in Table~\ref{ResPV}  but with the different bachelor mesons $\bar{D}^{*0}$ and $D_{(s)}^{*-}$; 
              the results in column $\Gamma_L/\Gamma$ are the predictions for the corresponding longitudinal 
              polarization fraction.}
\setlength{\tabcolsep}{12pt}  
\begin{tabular}{l c c} \hline\hline
  \;\;Decay modes                                                                      
                                                 &$\mathcal{B}$ (in $10^{-3}$)\;            &$\Gamma_L/\Gamma$\;       \\  
             \hline
  $B^+  \to\bar{D}^{*0}[\rho(770)^+ \to] \omega\pi^+$\; 
                                                &\;$1.21^{+0.17+0.09+0.05+0.07}_{-0.17-0.09-0.03-0.07}$\; 
                                                &\;$0.74^{+0.02}_{-0.02}$ \\ 
  $B^+  \to\bar{D}^{*0}[\rho(1450)^+ \to] \omega\pi^+$\; 
                                                &\;$0.87^{+0.12+0.07+0.03+0.37}_{-0.12-0.07-0.02-0.37}$\; 
                                                 &\;$0.67^{+0.02}_{-0.02}$ \\                                                       
            \hline                                                                                
  $B^0   \to D^{*-}    [\rho(770)^+ \to] \omega\pi^+$\;  
                                                &\;$1.20^{+0.18+0.09+0.02+0.07}_{-0.18-0.08-0.01-0.07}$\;  
                                                 &\;$0.68^{+0.02}_{-0.02}$  \\ 
  $B^0   \to D^{*-}    [\rho(1450)^+ \to] \omega\pi^+$\;     
                                                &\;$0.89^{+0.13+0.06+0.02+0.38}_{-0.13-0.06-0.02-0.38}$\;  
                                                 &\;$0.63^{+0.01}_{-0.01}$  \\        
            \hline
  $B_s^0 \to D_s^{*-}  [\rho(770)^+ \to] \pi\pi^+$\;  
                                                &\;$1.03^{+0.11+0.08+0.00+0.05}_{-0.11-0.08-0.00-0.05}$\;  
                                                 &\;$0.65^{+0.01}_{-0.01}$  \\    
  $B_s^0 \to D_s^{*-}  [\rho(1450)^+ \to] \pi\pi^+$\;  
                                                &\;$0.77^{+0.08+0.06+0.00+0.32}_{-0.08-0.06-0.00-0.32}$\;  
                                                 &\;$0.59^{+0.01}_{-0.01}$ \\                                                                                                
\hline\hline
\end{tabular}\label{ResVV}   
\end{table*}

Utilizing differential branching fractions the Eq.~(\ref{eqn-diff-bra}) and the decay amplitudes collected in Appendix, 
we obtain the branching fractions in Tables~\ref{ResPV}-\ref{ResVV} for the concerned quasi-two-body decays with 
$\rho(770)^+$ and $(1450)^+$ decaying into $\omega\pi^+$. For these PQCD branching fractions in 
Tables~\ref{ResPV}-\ref{ResVV}, their first error comes from the uncertainties of the shape parameter 
$\omega_B=0.40\pm0.04$ or $\omega_{B_s}=0.50 \pm 0.05$ for the $B^{\pm,0}$ or $B^0_s$ meson; 
the Gegenbauer moments $C_D=0.6\pm0.15$ or  $C_{D^*}=0.5\pm0.10$ for $D_{(s)}$ or $D^*_{(s)}$ mesons 
contribute the second error; the third one is induced by the Gegenbauer moments $a_R^{0}=0.25\pm0.10$, 
$a_R^t=-0.60\pm0.20$ and $a_R^s=0.75\pm0.25$~\cite{plb763-29} for the intermediate states; the fourth one  
for the decay results with $\rho\to \omega\pi$ comes from the uncertainties of the coupling 
$g_{\rho\omega\pi}$ or $A_1$ in Eq.~(\ref{exp-formfactor}).   
The uncertainties of the PQCD results in Table~\ref{tab_Rpp} are obtained by adding the individual theoretical errors 
in quadrature which induced by the uncertainties of $\omega_{B_{(s)}}$, $C_{D^{(*)}}$ and $a_R^{0,t,s}$, respectively.
There are other errors for the PQCD predictions in this work, which come from the uncertainties of the masses and 
the decay constants of the initial and final states, from the uncertainties of the Wolfenstein parameters, etc., are small 
and have been neglected.

\begin{table}[thb] 
\begin{center}                                                                   
   \caption{Experimental data for the relevant three-body branching fractions from 
                 {\it Review of Particle Physics}~\cite{PDG22}.}
\label{tab-data}   
\setlength{\tabcolsep}{15pt}  
\begin{tabular}{l c} \hline\hline
     \; Decay mode                                & \;\;${\mathcal B}$~\cite{PDG22}        \\         \hline             
     $B^+ \to \bar{D}^0\omega\pi^+$\;         &\;$(4.1\pm0.9)\times 10^{-3}$              \\ 
     $B^+ \to \bar{D}^{*0}\omega\pi^+$\;     &\;$(4.5\pm1.2)\times 10^{-3}$               \\       
     $B^0 \to        {D}^-\omega\pi^+$\;         &\;$(2.8\pm0.6)\times 10^{-3}$               \\     
     $B^0 \to        {D}^{*-}\omega\pi^+$\;     &\;$(2.8\pm0.6)\times 10^{-3}$                \\             
\hline\hline
\end{tabular}   
\end{center}
\end{table}

The four decay channels $B^+ \to \bar{D}^{(*)0} \omega\pi^+$ and ${B}^0 \to {D}^{(*)-} \omega\pi^+$ have been 
observed by CLEO Collaboration in Ref.~\cite{prd64.092001},  the updated studies for the decay 
${B}^0 \to {D}^{*-} \omega\pi^+$ were presented later by {\it BABAR} and Belle Collaborations in 
Refs.~\cite{prd74.012001,prd92.012013}. In these measurements, the $\omega\pi^+$ system in the final states 
showed a preference for the $1^-$ resonances. The relevant data from {\it Review of Particle Physics}~\cite{PDG22} 
are found in Table~\ref{tab-data}. In addition to the total branching fraction for ${B}^0 \to {D}^{*-} \omega\pi^+$ decay, 
one finds the fitted branching fractions 
\begin{eqnarray}
   \mathcal{B}&=&(1.48\pm0.27^{+0.15+0.21}_{-0.09-0.56})\times 10^{-3}\\
   \mathcal{B}&=&(1.07^{+0.15+0.06+0.40}_{-0.31-0.13-0.02})\times 10^{-3}
\end{eqnarray}
in Ref.~\cite{prd92.012013} for the 
quasi-two-body decays ${B}^0 \to{D}^{*-} \rho(770)^+\to {D}^{*-} \omega\pi^+$ and 
${B}^0 \to {D}^{*-} \rho(1450)^+\to {D}^{*-} \omega\pi^+$, respectively, where the first error is statistical, 
the second is systematic and the third is the model error. One can find that the predictions 
\begin{eqnarray}
 \mathcal{B}(B^0 \to D^{*-} [\rho(770)^+ \to] \omega\pi^+) &=&
         (1.20^{+0.18+0.09+0.02+0.07}_{-0.18-0.08-0.01-0.07})\times 10^{-3}, \\        
  \mathcal{B}(B^0 \to D^{*-} [\rho(1450)^+ \to] \omega\pi^+)&=&
         (0.89^{+0.13+0.06+0.02+0.38}_{-0.13-0.06-0.02-0.38})\times 10^{-3}                   
\end{eqnarray}
in Table~\ref{ResVV} for the corresponding two quasi-two-body decays are in agreement with these two branching 
fractions presented by Belle Collaboration in~\cite{prd92.012013}. In consideration of the fitted branching fractions for 
${B}^0 \to  {D}^{*-} \rho(770)^+\to {D}^{*-} \omega\pi^+$ and 
${B}^0 \to  {D}^{*-} \rho(1450)^+\to {D}^{*-} \omega\pi^+$ in~\cite{prd92.012013} and the data  
in Table~\ref{tab-data} for the three-body decay $B^0 \to {D}^{*-}\omega\pi^+$, one finds that 
the contributions from subprocesses $\rho(770)^+\to \omega\pi^+$ and $\rho(1450)^+\to \omega\pi^+$ are 
dominant for this three-body process.

By examining the fraction of the longitudinal polarization $\Gamma_L/\Gamma$ at a fixed value of the 
momentum transfer, the decays $B^0 \to {D}^{*-}\rho(770,1450)^+ \to{D}^{*-}\omega\pi^+$ can be employed 
to test the factorization hypothesis for $B$ meson decays~\cite{plb89-105,prd67.112002}. 
The measurement of the fraction of longitudinal polarization in Ref.~\cite{prd67.112002} for the decays 
$B^0 \to {D}^{*-}\rho(770)^+$ and $B^+  \to\bar{D}^{*0}\rho(770)^+$ confirmed the validity of the factorization 
assumption at relatively low region of the momentum transfer. 
In Ref.~\cite{plb507-142}, the authors proposed that if the $\omega\pi^+$ system in the 
${B} \to {D}^{*} \omega\pi^+$ decays is composed of two or more particles not dominated by a single narrow resonance, factorization can be tested in different kinematic regions. In Table~\ref{ResVV}, we list PQCD predictions for the 
corresponding longitudinal polarization fractions $\Gamma_L/\Gamma$ for the relevant decays. The errors, which are 
added in quadrature, for these longitudinal polarization fractions are quite small from the uncertainties of 
$\omega_{B_{(s)}}$, $C_{D^{(*)}}$, Gegenbauer moments for resonances, coupling $g_{\rho\omega\pi}$ 
or the weight parameter $A_1$.  The explanation is that the increase or decrease for the relevant numerical 
results from the uncertainties of these parameters will result in nearly identical change of the weight for the 
numerator and denominator of the corresponding $\Gamma_L/\Gamma$ predictions. 
In Ref.~\cite{prd64.092001}, the longitudinal polarization fraction for $B^0 \to {D}^{*-}\omega\pi^+$ was measured to be 
$\Gamma_L/\Gamma=0.63\pm0.09$; for the same decay channel in mass region of $1.1$-$1.9$ GeV for $\omega\pi^+$,  the result $\Gamma_L/\Gamma=0.654\pm0.042(\rm stat.)\pm0.016(\rm syst.)$ was provided by {\it BABAR} in 
Ref.~\cite{prd74.012001}. These two measurements agree well with the corresponding predictions in Table~\ref{ResVV}.

\begin{figure}[tbp]  
\centerline{\epsfxsize=8cm \epsffile{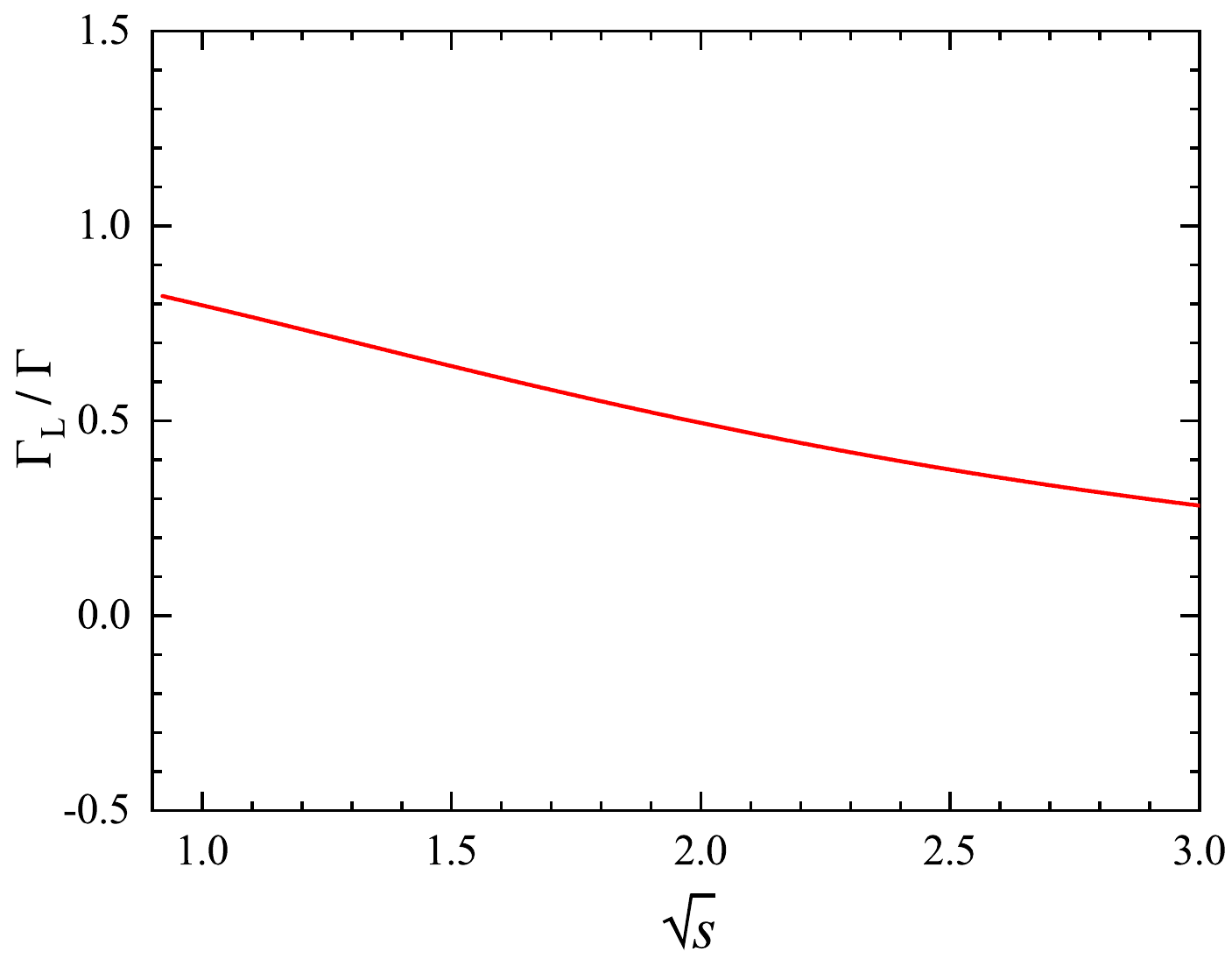}}
\caption{The invariant mass $\sqrt s$ dependent $\Gamma_L/\Gamma$ for $B_s^0 \to D_s^{*-}\rho(770)^+$ decay, 
              with the subprocess $\rho(770)^+ \to \omega\pi^+$. }
\label{FigGL}
\vspace{-0.2cm}
\end{figure}

When employing $B^0 \to {D}^{*-}\rho(770,1450)^+ \to{D}^{*-}\omega\pi^+$ to test of the factorization hypothesis, 
we should keep in mind that there are contributions from the annihilation Feynman diagrams as shown in 
Fig.~\ref{fig-feyndiag}-(c) for these two decay processes. By comparing the data $\mathcal{B}=(3.2^{+1.5}_{-1.3}) 
\times 10^{-5}$ for the pure annihilation decay $B^0\to D_s^{*-} K^{*+}$~\cite{PDG22} with the results in 
Table~\ref{tab_Rpp} for $B^0 \to {D}^{*-}\rho(770)^+$, one can roughly take the annihilation diagram contributions 
to be around a few percent at the decay amplitude level. In order to avoid the pollution from annihilation Feynman 
diagrams, we recommend to take the decays $B_s^0 \to D_s^{*-}\rho(770,1450)^+$ with $\rho(770,1450)^+$ decay into 
$\pi^+\pi^0$ or $\omega\pi^+$ to test of the factorization hypothesis, in view of these decay channels have only 
emission diagrams the Fig.~\ref{fig-feyndiag}-(b) at quark level. We plot the invariant mass $\sqrt s$ dependent 
$\Gamma_L/\Gamma$ in Fig.~\ref{FigGL} for the decay $B_s^0 \to D_s^{*-}\rho(770)^+$ with the subprocess 
$\rho(770)^+ \to \omega\pi^+$. One finds that the $\Gamma_L/\Gamma$ for $B_s^0 \to D_s^{*-}\rho(770)^+$
is going down as the increase of the invariant mass $\sqrt{s}$ for $\omega\pi^+$ system.  
Since the subprocesses $\rho(770)^+ \to \pi^+\pi^0$ and $\rho(770)^+ \to \omega\pi^+$ are described by the 
electromagnetic form factors $F_\pi$ and $F_{\omega\pi}$, respectively, in the quasi-two-body decay amplitudes, 
they are independence from the weak interaction in the related decay processes and wouldn't disturb the 
measurement results of $\Gamma_L/\Gamma$ for relevant channels.

\begin{figure}[tbp]  
\centerline{\epsfxsize=7.5cm \epsffile{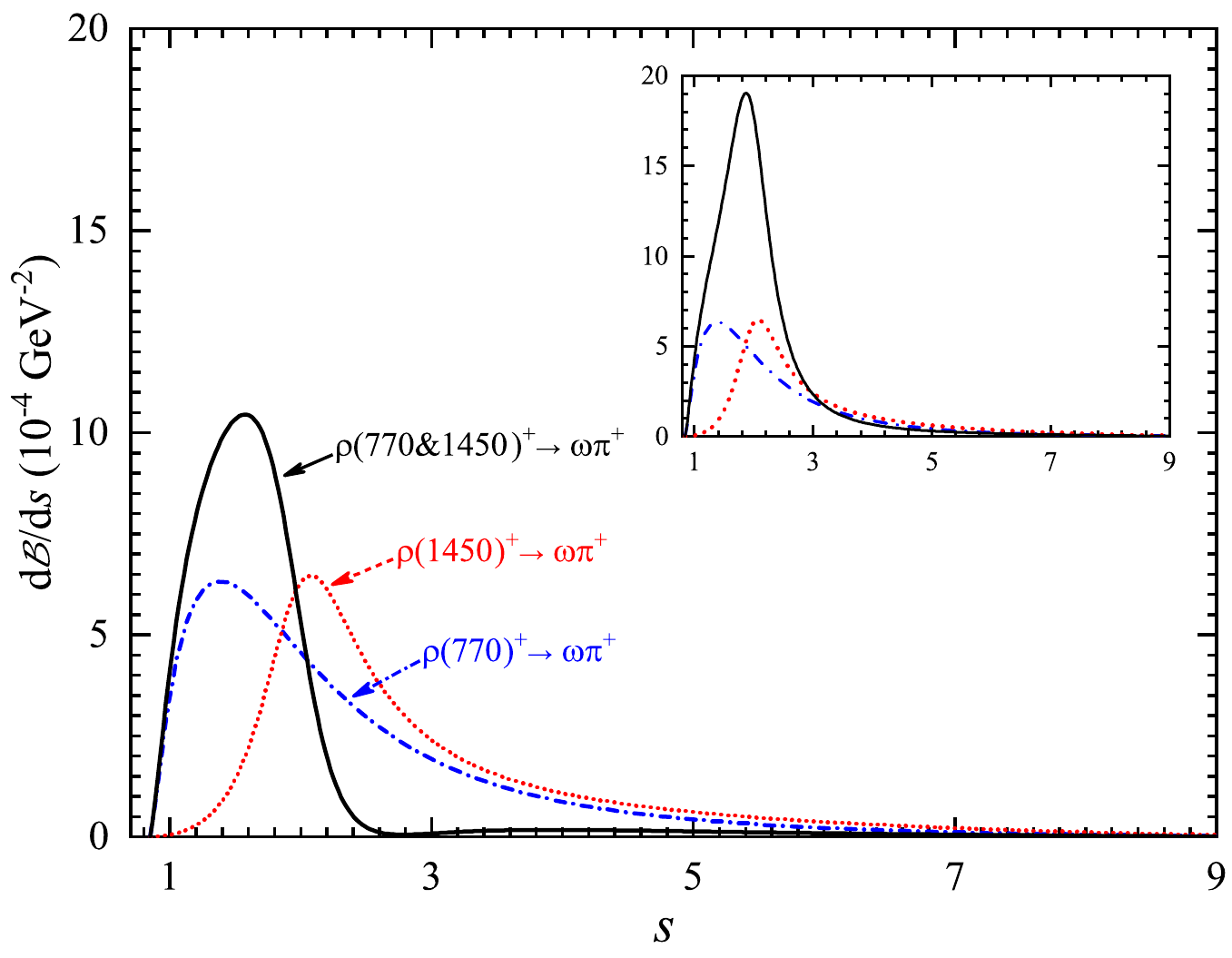} }
\centerline{ \epsfxsize=8cm \epsffile{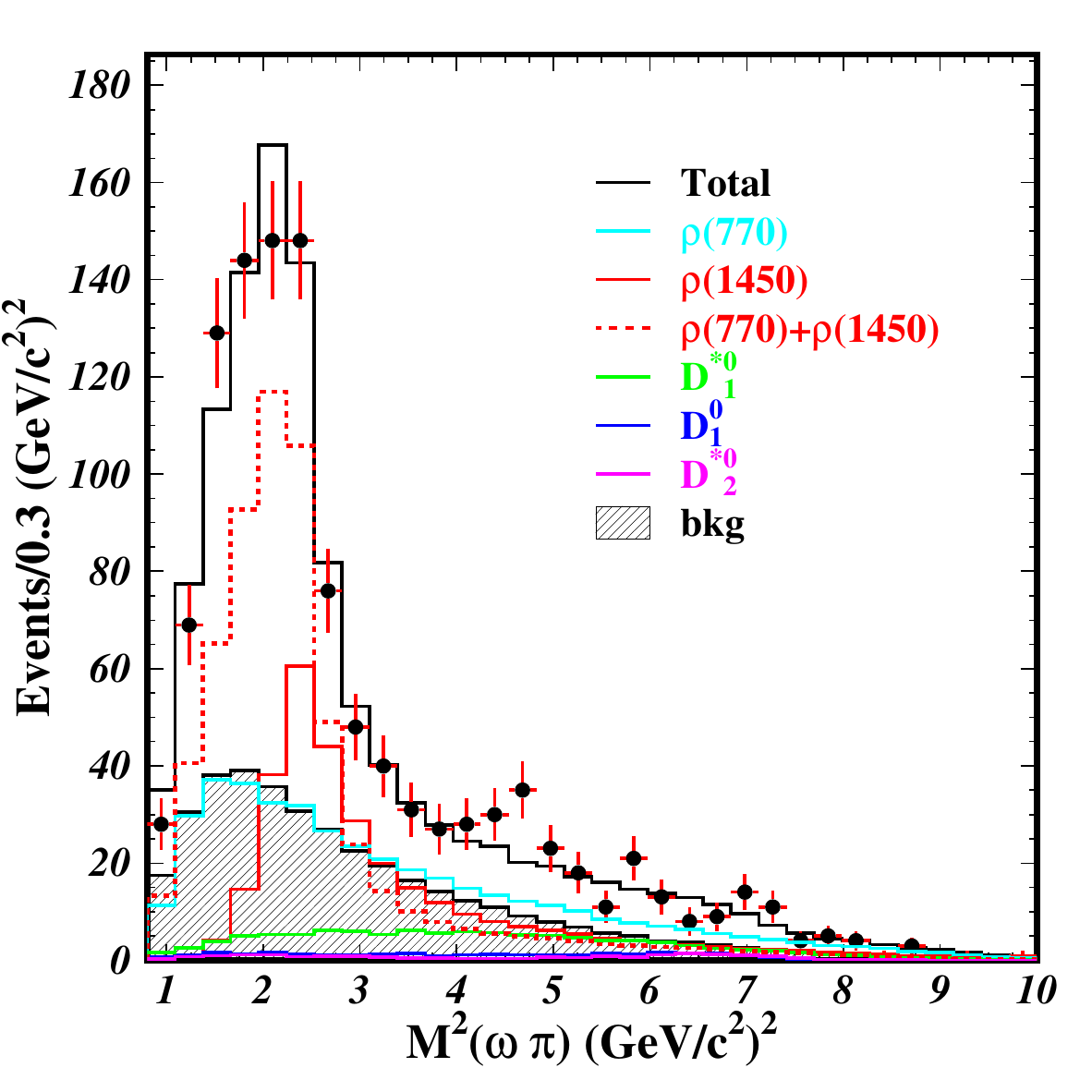} }
\caption{The predicted differential branching fraction (top diagram) 
                 for the quasi-two-body decay $B^0 \to {D}^{*-}\rho^+\to {D}^{*-}\omega\pi^+$, 
                 the inset is for the phase difference $\phi_1=0.6\pi$.
                 Along with the distribution for $\omega\pi$ (bottom one) 
                 for  $B^0 \to  {D}^{*-}\omega\pi^+$ measured by Belle Collaboration in~\cite{prd92.012013}.
                  }
\label{FigRes}
\vspace{-0.2cm}
\end{figure}

In the top diagram of Fig.~\ref{FigRes}, we show the differential branching fraction for the quasi-two-body decays
$B^0\to {D}^{*-}\rho^+\to{D}^{*-}\omega\pi^+$,  $\rho^+\in \{\rho(770)^+,\rho(1450)^+,\rho(770)^+\&\rho(1450)^+\}$. 
The phase difference $\phi_1$ between $\rho(770)$ and $\rho(1450)$ for Eq.~(\ref{exp-formfactor}) will generate 
different shapes for the curves of differential branching fractions and branching fractions of the decay processes with 
$\rho(770\&1450)^+\to\omega\pi^+$. In Refs.~\cite{plb486-29,plb562-173,JETPL94-734,prd61.072003}, a phase 
difference of $\phi_1=\pi$ were adopted or fitted for $F_{\omega\pi}(s)$ between $\rho(770)$ and $\rho(1450)$; 
the measurements in~\cite{prd88.054013,prd94.112001,prd92.012013} showing the results close to 
$\pi$ for this phase $\phi_1$.  With the choice of phase $\phi_1=\pi$  we find the shapes of these curves 
in the top diagram of Fig.~\ref{FigRes} doesn't agree very well with the distribution of 
$\omega\pi$ for  $B^0 \to  {D}^{*-}\omega\pi^+$ decay measured by Belle Collaboration in~\cite{prd92.012013} 
and shown in the bottom diagram of Fig.~\ref{FigRes}. We find the curve for $B^0\to {D}^{*-}[\rho(770)^+\&\rho(1450)^+\to]
\omega\pi^+$ is seriously affected by the interference between $\rho(770)^+$ and $\rho(1450)^+$ from the top diagram 
of Fig.~\ref{FigRes}. Since the phase difference of $\phi_1=\pi$, there is essentially a minus sign 
between $\rho(770)$ and $\rho(1450)$ components of  Eq.~(\ref{exp-formfactor}) the form factor $F_{\omega\pi}(s)$.  
Take into consideration of the denominator $D_{\rho_i}$ for BW formula, in the invariant mass region of $\omega\pi^+$ 
system well below the mass of $\rho(1450)$, the form factor $F_{\omega\pi}(s)$ will be instructive for the branching 
fraction of $B^0\to {D}^{*-}[\rho(770)^+\&\rho(1450)^+\to]\omega\pi^+$, but when invariant mass 
is much larger than the mass for $\rho(1450)$, the real parts of denominator $D_{\rho_i}$ for $\rho(770)^+$ and 
$\rho(1450)^+$ will have the same sign, the $F_{\omega\pi}(s)$ will be destructive even if we take the influence  
of the full width of $\rho(1450)$ into account.  We alter $\phi_1$ from zero to $2\pi$, and find the predicted curve 
for $d{\mathcal{B}}/d s$ will match Belle's figure better when we employ $\phi_1=0.6\pi$ as shown in the inset of 
Fig.~\ref{FigRes}~(top). It also illustrates the phase difference between $\rho(770)$ and $\rho(1450)$ for $F_{\omega\pi}(s)$ 
could be different in the electromagnetic form factor and $B$ decays.  The branching fractions with the subprocess 
$\rho(770\&1450)^+\to\omega\pi^+$ could also verify our analysis above. For example, we have the prediction 
\begin{eqnarray} %
  \mathcal{B} =(1.03^{+0.16+0.07+0.01+0.15}_{-0.16-0.08-0.01-0.15})\times 10^{-3}
\end{eqnarray}
for $B^0 \to D^{*-}  [\rho(770\&1450)^+ \to] \omega\pi^+$ decay when $\phi_1=\pi$, 
it is much smaller than the sum of two branching fractions from the subprocesses $\rho(770)^+\to\omega\pi^+$ 
and $\rho(1450)^+\to\omega\pi^+$ given in Table~\ref{ResVV}. 

Because the threshold for $\omega\pi^+$ is larger than the mass of $\rho(770)^+$, we don't see a typical BW shape 
for the curve with subprocess $\rho(770)^+\to\omega\pi^+$ in Fig.~\ref{FigRes}, the bump of the curve is attributed to 
the kinematic characteristics in corresponding decay process rather than the properties of the involved resonant state 
$\rho(770)^+$. The resonance $\rho(770)^+$ as a virtual bound state~\cite{Dalitz62,plb25-294} 
in the process $\rho(770)^+\to\omega\pi^+$ can not completely present its properties in the concerned processes 
because of the phase space of the relevant decay processes. But the quantum number of the involved resonance could 
be fixed from its decay daughters the $\omega\pi^+$ system. The exact resonant source for $\omega\pi^+$ makes the 
cascade decay like $B^0\to {D}^{*-}\rho(770)^+\to{D}^{*-}\omega\pi^+$ to be a quasi-two-body process, although the  
invariant mass region for the $\omega\pi^+$ system is excluded from the region around pole mass of $\rho(770)$.  
The resonance $\rho(1450)^+$ with the mass larger than the threshold of $\omega\pi^+$ contribute a normal BW shape 
for the curve of the differential branching fraction in Fig.~\ref{FigRes} for the decay $B^0\to {D}^{*-}\rho(1450)^+ 
\to{D}^{*-}\omega\pi^+$. But in the decay $D^+_s\to \omega\pi^+\eta$ which has been measured by 
BESIII recently~\cite{prd107.052010}, since the initial decaying state $D^+_s$ does not have enough energy to make 
$\rho(1450)$ demonstrate its intact properties, it will provide only the virtual contribution for $\omega\pi^+$ 
system in this three-body $D^+_s$ decay process, we shall leave the detailed discussion of it to future study.

\section{Summary}
\label{sec-sum}
In this work we studied the subprocesses $\rho(770,1450)^+\to \omega\pi^+$ contributions in the cascade decays 
$B^{+}\to \bar{D}^{(*)0} \rho^+ \to \bar{D}^{(*)0}\omega\pi^+$, $B^{0}\to D^{(*)-} \rho^+ \to D^{(*)-}\omega\pi^+$ 
and $B_s^{0}\to D_s^{(*)-} \rho^+ \to D^{(*)-}\omega\pi^+$ within the PQCD approach. These decays are 
important for the investigation of the properties for  $\rho$ excitations and are very valuable for the test of 
the factorization hypothesis for $B$ meson decays. The decays $B^{+}\to \bar{D}^{(*)0}\omega\pi^+$ and 
$B^{0}\to D^{(*)-}\omega\pi^+$ have been measured by different collaborations but without any predictions 
for their observables on theoretical side.

With one open charm meson in the final state of each decay channel, the decay amplitudes of these processes 
were described well by effective Hamiltonian $\mathcal{H}_{\rm eff}$ with the tree-level $W$ exchange operators 
$O_1$ and $O_2$ in the quasi-two-body framework. The subprocesses $\rho(770,1450)^+\to \omega\pi^+$, 
which are related to the processes $e^+e^- \to \omega\pi^0$ and $\tau \to \omega \pi \nu_\tau$ and can not 
be calculated in PQCD, were introduced into the distribution amplitudes for $\omega\pi$ system in this work 
via the vector form factor $F_{\omega\pi}(s)$ which has measured by different collaborations recently.

With the parameters $g_{\rho\omega\pi}=16.0\pm2.0$ GeV$^{-1}$ and $A_1=0.171\pm0.036$ for form factor 
$F_{\omega\pi}(s)$, we predicted the branching fractions for the first time on theoretical side for $12$ quasi-two-body 
decays with $\rho(770,1450)^+\to \omega\pi^+$, as well as the corresponding longitudinal polarization fractions 
$\Gamma_L/\Gamma$ for the cases with the vector $\bar{D}^{(*)0}$ or $D_{(s)}^{(*)-}$ in their final states.
The branching fractions of these quasi-two-body decays are at the order of $10^{-3}$, which can be detected at 
the LHCb and Belle-II experiments. Our results 
$\mathcal{B}=(1.20^{+0.18+0.09+0.02+0.07}_{-0.18-0.08-0.01-0.07})\times 10^{-3}$ and 
$\mathcal{B}=(0.89^{+0.13+0.06+0.02+0.38}_{-0.13-0.06-0.02-0.38})\times 10^{-3}$ for the decays               
${B}^0 \to{D}^{*-} \rho(770)^+\to {D}^{*-} \omega\pi^+$ and ${B}^0 \to {D}^{*-} \rho(1450)^+\to {D}^{*-} \omega\pi^+$
agree with the measurements $\mathcal{B}=(1.48\pm0.27^{+0.15+0.21}_{-0.09-0.56})\times 10^{-3}$ and
$\mathcal{B}=(1.07^{+0.15+0.06+0.40}_{-0.31-0.13-0.02})\times 10^{-3}$, respectively, from Belle  Collaboration.

The decay $B^0\to{D}^{*-}\omega\pi^+$ has been employed in literature to test the factorization hypothesis for 
$B$ meson decays by examining the fraction of the longitudinal polarization $\Gamma_L/\Gamma$ at a fixed value 
of the momentum transfer. But we should take care about contributions from the annihilation Feynman diagrams for
this decay process. In order to avoid the pollution from annihilation Feynman diagrams, we recommend to take the decays 
$B_s^0 \to D_s^{*-}\rho(770,1450)^+$ with $\rho(770,1450)^+$ decay into $\pi^+\pi^0$ or $\omega\pi^+$ to test 
the factorization hypothesis for $B$ decays. These decay channels have only emission diagrams with $B_s\to D_s^{*-}$ 
transition at quark level, and the subprocesses which can be described with the corresponding electromagnetic form 
factors would not disturb the measurement results for $\Gamma_L/\Gamma$.
  
The resonance $\rho(770)^+$ in the concerned quasi-two-decays of this work decaying to $\omega\pi^+$ system in the final states can not completely present its properties and contribute only the virtual contribution for the total branching 
fraction for corresponding three-body decay channels, because of the threshold for $\omega\pi^+$ and 
phase space limitation. But the quantum number of the involved resonance could be fixed from its decay daughters 
the $\omega\pi^+$ system. We want to stress here that the virtual contributions from specific known intermediate states 
are different from the nonresonant contributions demarcated in the experimental studies.

\acknowledgments
W. F. Wang thanks Prof.~\`Angels Ramos for helpful discussions and for carefully reading the manuscript;
we also thank Prof. Cai-Dian L\"{u} for valuable discussions. 
This work was supported in part by the National Natural Science Foundation of China under Grants 
No. 12205148 and No. 11947011,  the Fund for Shanxi ``1331 Project" Key Subjects Construction, 
the Natural Science Foundation of Jiangsu Province under Grant No. BK20191010, 
and the Qing Lan Project of Jiangsu Province.

\appendix
\section{Decay amplitudes for $B\to\bar{D}^{(*)}\rho \to \bar{D}^{(*)}\omega\pi$ decays}  
\label{sec-appx}

With the effective weak Hamiltonian $\mathcal{H}_{\rm eff}$ in Eq.~(\ref{eff_Ham}), 
the total decay amplitudes for the concerned quasi-two-body decays are then written as
\begin{eqnarray}
  & &  \mathcal{A}(B^+ \to \bar{D}^0 [\rho^+ \to] \omega\pi^+)=\frac{G_F}{\sqrt2}V^*_{cb}V_{ud}   %
  \big[a_1F_{T\rho}^{LL} +C_2 M_{T\rho}^{LL}+a_2F_{TD}^{LL}+C_1M_{TD}^{LL} \big],  \quad  \label{DA_Bu}    \\
  & & \mathcal{A}(B^0 \to {D^-} [\rho^+ \to] \omega\pi^+)=\frac{G_F}{\sqrt2}V^*_{cb}V_{ud}  %
    \big[a_2F_{TD}^{LL}+C_1M_{TD}^{LL} +a_1F_{a\rho}^{LL}  + C_2 M_{a\rho}^{LL} \big],  \quad   
\label{DA_B0} \\
  & & \mathcal{A}(B_s^0 \to D_s^- [\rho^+ \to] \omega\pi^+)=\frac{G_F}{\sqrt2}V^*_{cb}V_{ud} \big[a_2F_{TD}^{LL}      
          +C_1M_{TD}^{LL} \big],
 \label{DA_Bs0}
\end{eqnarray}
by combining various of contributions from the related Feynman diagrams in Fig.~\ref{fig-feyndiag}.  
Where $\rho^+$ stands for the $\rho(770)^+$ or $\rho(1450)^+$ in the relevant decays. The other three decay 
amplitudes for the corresponding $B^+$, $B^0$ and $B_s^0$ decays with $\bar{D}^{*0}$, $D^{*-}$ and $D_s^{*-}$, 
respectively, can be obtained from Eq.~(\ref{DA_Bu})-(\ref{DA_Bs0}) with the replacements of $D_{(s)}$ meson wave 
function by the $D^*_{(s)}$ wave function. As has been done in two-body decays of $B$ to two vector mesons as the 
final state, the decay amplitudes for $B \to \bar{D}^*\rho^+\to \bar{D}^*\omega\pi^+$ in this work can be 
decomposed as $\mathcal{A}^{(\lambda)}=M^{(\lambda)}\cdot{\langle \omega\pi\vert \rho_i \rangle}/{D_{\rho_i}(s)}$
with~\cite{prd76.074018}
\begin{eqnarray}
M^{(\lambda)}
 &=&\epsilon_{\bar{D}^*\mu}^{*}(\lambda)\epsilon_{\rho\nu}^{*}(\lambda) \left[ a
      \,g^{\mu\nu} + {b \over m_{D} \sqrt{s} } P_{B}^\mu P_{B}^\nu + i{c \over m_{D}\sqrt{s}} 
      \epsilon^{\mu\nu\alpha\beta} P_{\alpha} P_{3\beta}\right]\!,
  \nonumber \\
&\equiv &M_{L}+M_{N}
     \epsilon^{*}_{\bar{D}^*}(\lambda=T)\cdot\epsilon^{*}_{\rho}(\lambda=T) +i \frac{M_{T}}{m_{B}^2}
     \epsilon^{\alpha \beta\gamma \rho} \epsilon^{*}_{\rho\alpha}(\lambda)\epsilon^{*}_{D^*\beta}(\lambda) 
     P_{\gamma }P_{3\rho }. \quad
\end{eqnarray}
According to the polarized decay amplitudes, one has $|A|^2=|A_L|^2+|A_\parallel|^2+|A_\perp|^2$, and 
$\Gamma_L/\Gamma=|A_L|^2/(|A_L|^2+|A_\parallel|^2+|A_\perp|^2)$, the amplitudes $A_L, A_\parallel$ and 
$A_\perp$ are related to the $M_{L}, M_{N}$ and $M_{T}$, respectively. For the detailed discussion, one is 
referred to Refs.~\cite{prd76.074018,prd45.193,prd46.2969,zpc55-497,prd61.074031}.

With the subprocesses $\rho^+\to \omega\pi^+$, where $\rho$ is $\rho(770)$ or $\rho(1450)$, the specific expressions 
in PQCD approach for the Lorentz invariant decay amplitudes of these general amplitudes $F$'s and $M$'s for 
$B \to \bar{D}^{(*)} \rho\to\bar{D}^{(*)} \omega\pi$ decays are given as follows:

The amplitudes from Fig.~\ref{fig-feyndiag}-(a) for the decays with a pseudoscalar $\bar{D}^0$ or $D^-_{(s)}$ meson 
in the final state are written as
\begin{eqnarray}
F_{T\rho}^{LL}&=&8\pi C_F m^4_B f_{D}\int dx_B dx\int b_B db_B b db \phi_B \big\{\big[[r^2-\bar{\zeta}(x(r^2-1)^2+1)]\phi^0-\sqrt{\zeta}[(r^2+\bar{\zeta}\nonumber\\
&& +2\bar{\zeta}x(r^2-1))\phi^s -(r^2-1)\bar{\zeta} (2 x(r^2-1) +1)-r^2)\phi^t\big] E_e(t_a)h_a(x_B,x,b,b_B)S_t(x) \nonumber\\
&&+\big[(r^2-1)[\zeta \bar{\zeta}-r^2 (\zeta-x_B)]\phi^0-2 \sqrt{\zeta}[\bar{\zeta}-r^2 (x_B-2 \zeta+1)]\phi^s \big]\nonumber\\
&&\times E_e(t_b)h_b(x_B,x,b_B,b)S_t(|x_B-\zeta|)\big\},
\label{eq:f01}
\\
M_{T\rho}^{LL}&=&16\sqrt{\frac{2}{3}} \pi C_F m_B^4 \int dx_B dx dx_3 \int b_B db_B b_3 db_3 \phi_B \phi_D\big\{\big[
-[(\bar{\zeta}+r^2) ((r^2-1) (x_3 \bar{\zeta}+x_B) \nonumber\\
&&+r^2 (\zeta x-1)-\zeta (x+1)+1)+r r_c(r^2-\bar{\zeta})]\phi^0-\sqrt{\zeta}[(r^2 (\bar{\zeta} (x_3+x-2)+x_B)-x\bar{\zeta}\nonumber\\
&&+4 r r_c)\phi^s +(r^2-1)(r^2 (\bar{\zeta}(x-x_3)-x_B)-x \bar{\zeta}) \phi^t ]\big]E_n(t_c)h_c(x_B,x,x_3,b_B,b_3)\nonumber\\
&&+\big[x(r^2-1)[(r^2-\bar{\zeta})\phi^0 +\sqrt{\zeta} \bar{\zeta} (\phi^s-(r^2-1) \phi^t)]-(x_3 \bar{\zeta}-x_B)[(r^2-\bar{\zeta})\phi^0 \nonumber\\
&&+\sqrt{\zeta} r^2 ((r^2-1) \phi^t+\phi^s)]\big] E_n(t_d)h_d(x_B,x,x_3,b_B,b_3) \big\},
\label{eq:m01}
\end{eqnarray}
with the symbol $\bar{\zeta} = 1-\zeta$, the mass ratios $r=m_{D^{(*)}}/m_B$ and $r_c=m_c/m_B$.
The amplitudes from Fig.~\ref{fig-feyndiag}-(b) are written as
\begin{eqnarray}
F_{TD}^{LL}&=&8\pi C_F m^4_B f_\rho \int dx_B dx_3\int b_B db_B b_3 db_3 \phi_B \phi_D \big\{\big[
(r+1)[r^2-\bar{\zeta}-x_3 \bar{\zeta}(r-1) (2 r-\bar{\zeta})]\big] \nonumber\\
&&\times E_e(t_m)h_m(x_B,x_3,b_3,b_B)S_t(x_3)+\big[(r^2-\bar{\zeta})[2 r (r_c+1)-r^2 \bar{\zeta}-r_c]-\zeta x_B (2 r-\bar{\zeta}) \big] \nonumber\\
&&\times E_e(t_n)h_n(x_B,x_3,b_B,b_3)S_t(x_B) \big\},
\label{eq:f03}
\\
M_{TD}^{L L} &=& 16\sqrt{\frac{2}{3}} \pi C_F m_B^4\int d x_B d x d x_3 \int b_B d b_B b d b \phi_B \phi_D \phi^0 \big\{\big[ x_B [\bar{\zeta}^2-\bar{\zeta}r^2+\zeta r]+\bar{\zeta}x_3 r (\zeta r \nonumber\\
&&+(r+1) (r-1)^2)-\zeta (r-1)^2 (r+1)[(r+2) x-2 (r+1)]+\zeta^2 [x-r^2(x-2)-1]  \nonumber\\
&&+(x-1)(r^2-1)^2\big] E_n(t_o) h_o(x_B, x, x_3, b_B, b)+\big[(r-1) (\bar{\zeta}+r) [x_B+(r^2-1) x]\nonumber\\
&&+\bar{\zeta} x_3[(r-1)^2 (r+1)-\zeta] \big] E_n(t_p) h_p(x_B, x, x_3, b_B, b)\big\}.
\label{eq:m03}
\end{eqnarray}
The amplitudes from Fig.~\ref{fig-feyndiag}-(c) the annihilation diagrams are written as
\begin{eqnarray}
 F_{A\rho}^{LL}&=&8\pi C_F m^4_B f_B\int dx_3dx\int bdbb_3db_3\phi_D \big\{\big[ 
((2r r_c-1) (r^2-\bar{\zeta})-(r^2-1)^2 x \bar{\zeta})\phi^0+\sqrt{\zeta} \nonumber\\
&&\times[(r^2-1)(r_c (r^2-\bar{\zeta})-2 r(r^2-1) x) \phi^t+(r_c (r^2-\zeta+1)+2 r (x-x r^2-2))\phi^s] \big]\nonumber\\
&&\times E_a(t_e)h_e(x,x_3,b,b_3)S_t(x)+\big[ (r^2-1)[x_3 \bar{\zeta}^2-\zeta (r^2-\bar{\zeta})]\phi^0+2 \sqrt{\zeta}r(x_3 \bar{\zeta}+\zeta\nonumber\\
&&-r^2+1)\phi^s \big]E_a(t_f)h_f(x,x_3,b_3,b) S_t(|\bar{\zeta}x_3+\zeta|)\big\},
\label{eq:f02}
\\
M_{A\rho}^{LL}&=&-16\sqrt{\frac{2}{3}} \pi C_F m_B^4\int dx_B dx dx_3\int b_Bdb_Bbdb\phi_B\phi_D\big\{\big[
(r^2-1)[r^2 (x_B+x_3-1)+x_B \nonumber\\
&&+x_3] \phi^0+\zeta [r^4 x-(r^2-1) x_B+\zeta ((r^2-1) x_3-x r^2+x+1)-(r^4+r^2-2) x_3\nonumber\\
&&-x-1]\phi^0 +\zeta^{3/2}r(1-x_3)[(r^2-1)\phi^t+\phi^s]+\sqrt{\zeta}r[\phi^s (x_B+r^2 (x-1)+x_3-x+3)\nonumber\\
&&+(r^2-1)(x_B-x r^2+r^2+x_3+x-1)\phi^t] \big] E_n(t_g)h_g(x_B,x,x_3,b,b_B)+\big[(r^2-\bar{\zeta})\nonumber\\
&&\times [r^2 (x_B-x_3-x+1)+\zeta (r^2 (x_3+x-2)-x+1)+x-1]\phi^0 +\sqrt{\zeta}r[(x_B-x_3 \bar{\zeta}\nonumber\\
&&-\zeta+(r^2-1) (1-x))\phi^s+(1-r^2)(x_B-x_3 \bar{\zeta}-\zeta+(r^2-1) (x-1))\phi^t] \big]\nonumber\\
&&\times E_n(t_h)h_h(x_B,x,x_3,b,b_B)\big\}.
\label{eq:m02}
\end{eqnarray}
Where the $T\rho$, $TD$ and $A\rho$ in the subscript of above expressions stand for $B\to \rho$, $B\to D$ 
transitions and the annihilation Feynman diagrams, respectively. The $F$'s stand for those factorizable 
diagrams and $M$'s for the nonfactorizable diagrams in Fig.~\ref{fig-feyndiag}.

The longitudinal polarization amplitudes from Fig.~\ref{fig-feyndiag}-(a)  for the decays with a vector 
$\bar{D}^{*0}$ or $D^{*-}_{(s)}$ meson in the final state are written as
\begin{eqnarray}
F_{T\rho,L}^{LL}&=&8\pi C_F m^4_B f_{D^*}\int dx_B dx\int b_B db_B b db \phi_B \big\{\big[
[\bar{\zeta}+\bar{\zeta}x (r^2-1)^2 +(2 \zeta -1)r^2]\phi^0\nonumber\\
&&+\sqrt{\zeta} [(1-r^2) (2\bar{\zeta} x (r^2-1)+\bar{\zeta}+r^2)\phi^t+(2\bar{\zeta}x (r^2-1)+\bar{\zeta}-r^2)\phi^s]\big]
            \nonumber\\
&&\times E_e(t_a)h_a(x_B,x,b,b_B)S_t(x)+\big[(r^2-1) [r^2 x_B+\zeta ^2-\zeta (r^2+1)]\phi^0\nonumber\\
&&-2 \sqrt{\zeta} [r^2 (1-x_B)-\bar{\zeta}]\phi^s \big] E_e(t_b)h_b(x_B,x,b_B,b)S_t(|x_B-\zeta|)\big\},
\label{eq:f11}
\end{eqnarray}
\begin{eqnarray}
M_{T\rho,L}^{LL}&=&16\sqrt{\frac{2}{3}} \pi C_F m_B^4 \int dx_B dx dx_3 \int b_B db_B b_3 db_3 \phi_B \phi_{D^*}\big\{\big[[rr_c (1-\bar{\zeta}r^2-\zeta ^2)-(r^2-\bar{\zeta}) \nonumber\\
&&\times(\bar{\zeta} x_ 3 (r^2-1)+ x_B(r^2-1)+(\zeta  x-1)r^2 -\zeta (x+1)+1)]\phi^0-\sqrt{\zeta}[(r^2 (x_3 \bar{\zeta} -\bar{\zeta}x \nonumber\\
&&+x_B)-\zeta x+x)\phi^s +(r^2-1)(\bar{\zeta}x (1-r^2) -r^2 ((x_3-2) \bar{\zeta}+x_B))\phi^t ]\big]  \nonumber\\
&&\times E_n(t_c)h_c(x_B,x,x_3,b_B,b_3)+\big[x_B [(\bar{\zeta}+(2 \zeta -1) r^2)\phi^0 +\sqrt{\zeta} r^2 ((r^2-1) \phi^t+\phi^s)]\nonumber\\
&&-\bar{\zeta} x_ 3 [(\bar{\zeta}+(2 \zeta -1) r^2)\phi^0 +\sqrt{\zeta} r^2 ((r^2-1) \phi^t+\phi^s)]+x(r^2-1)[ (\bar{\zeta}+(2 \zeta -1) r^2)\phi^0  \nonumber\\
&&-\sqrt{\zeta} \bar{\zeta} (\phi^s-(r^2-1)\phi^t)]\big] E_n(t_d)h_d(x_B,x,x_3,b_B,b_3)\big\}.
\label{eq:m11}
\end{eqnarray}
The longitudinal polarization amplitudes from Fig.~\ref{fig-feyndiag}-(b) are 
\begin{eqnarray}
F_{TD^{*},L}^{LL}&=&8\pi C_F m^4_B f_\rho\int dx_B dx_3\int b_B db_B b_3 db_3\phi_B\phi_{D^*} \big\{\big[
\bar{\zeta}+(2 r-1) (r^2-1) x_3 \bar{\zeta}^2+r \nonumber\\
&&\times [\zeta(r^2+2r-\zeta)-r^2-r+1] \big]E_e(t_m)h_m(x_B,x_3,b_3,b_B)S_t(x_3)+\big[r^2[r_c(2 \zeta-1 ) \nonumber\\
&&-\zeta ^2+1]-\bar{\zeta} (\zeta x_B-r_c+r^4)\big] E_e(t_n)h_n(x_B,x_3,b_B,b_3)S_t(x_B)\big\},
\label{eq:f13}
\\
M_{TD^{*},L}^{LL} &=& -16\sqrt{\frac{2}{3}} \pi C_F m_B^4\int d x_B d x d x_3 \int b_B d b_B b d b \phi_B \phi_{D^*} \phi^0 \big\{\big[ \bar{\zeta} x_B (1-r)(\bar{\zeta}+r)-\bar{\zeta}x_3 r\nonumber\\
&& \times(r^3-\bar{\zeta}(r^2+r-1))-\zeta ^2 (2 r^3-x(r+1)(r-1)^2-2 r+1)+ (x-1)(r^2-1)^2\nonumber\\
&& -\zeta (r+1) (r-1)^2(r x+2x-2)\big]E_n(t_o)h_o(x_B,x,x_3,b_B,b)+\big[ \bar{\zeta} x_3[r^2 (r \bar{\zeta}+2 \zeta -1) \nonumber\\
&& +\bar{\zeta}-r \bar{\zeta}] - (x_B+(r^2-1) x)[\bar{\zeta}-\bar{\zeta}\zeta r+(2 \zeta -1) r^2]\big] E_n(t_p)h_p(x_B,x,x_3,b_B,b)\big\}.\nonumber\\
\label{eq:m13}
\end{eqnarray}
The longitudinal polarization amplitudes from Fig.~\ref{fig-feyndiag}-(c) are 
\begin{eqnarray}
 F_{A\rho,L}^{LL}&=&-8\pi C_F m^4_B f_B\int dx_3dx\int bdbb_3db_3\phi_{D^*} \big\{\big[
 \sqrt{\zeta} r_c [(r^4-\zeta r^2-\bar{\zeta})\phi^t+ (r^2-\bar{\zeta})\phi^s] \nonumber\\
&& +[\bar{\zeta}(1-x(r^2-1)^2 )+ r^2(2 \zeta -1)]\phi^0\big] E_a(t_e)h_e(x,x_3,b,b_3)S_t(x)+\big[2r\sqrt{\zeta}\bar{\zeta}
\nonumber\\
&&\times ((x_3-1) \bar{\zeta}+r^2) \phi^s+(r^2-1) [\zeta (r^2+\bar{\zeta}(1-x_3)-x_3)+x_3]\phi^0\big]\nonumber\\
&&\times E_a(t_f)h_f(x,x_3,b_3,b)S_t(|\bar{\zeta}x_3+\zeta|)\big\},
\label{eq:f12}
\end{eqnarray}
\begin{eqnarray}
M_{A\rho,L}^{LL}&=&16\sqrt{\frac{2}{3}} \pi C_F m_B^4\int dx_B dx dx_3\int b_Bdb_Bbdb\phi_B\phi_{D^*} \big\{\big[
 -(x_3\bar{\zeta}+x_B) [(r^2-1) (r^2-\bar{\zeta})\phi^0 \nonumber\\
&& -\sqrt{\zeta}\bar{\zeta} r (r^2-1) \phi^t-\sqrt{\zeta}\bar{\zeta} r \phi^s]-\zeta \bar{\zeta} (x+1)\phi^0+\sqrt{\zeta} \bar{\zeta} r^5 (x-1) \phi^t+\sqrt{\zeta}\bar{\zeta} r^3 ((\zeta-2 x)\phi^t \nonumber\\
&& -(x-1) \phi^s)+\sqrt{\zeta}\bar{\zeta} r((x-\bar{\zeta})\phi^s +(x+\bar{\zeta})\phi^t)-r^4(\zeta x-1) \phi^0 -r^2 (\zeta x(\zeta-2)+1)\phi^0 \big] \nonumber\\
&&\times E_n(t_g)h_g(x_B,x,x_3,b,b_B)-\big[(r^2-\bar{\zeta}) (r^2 (x_B-x_3 \bar{\zeta})+r^2 ( \bar{\zeta}x-1)- \bar{\zeta}(x-1))\phi^0   \nonumber\\
&& -\sqrt{\zeta}\bar{\zeta} r [((x_3-1)\bar{\zeta}+(1-x)r^2+x-x_B)\phi^s-(r^2-1)(x_3 \bar{\zeta}+\zeta-(1-x)r^2-x\nonumber\\
&& -x_B+1) \phi^t ]\big]E_n(t_h)h_h(x_B,x,x_3,b,b_B)\big\}.
\label{eq:m12}
\end{eqnarray}

The normal and transverse polarization amplitudes from Fig.~\ref{fig-feyndiag} for the decays with a vector 
$\bar{D}^{*0}$ or $D^{*-}_{(s)}$ are written as
\begin{eqnarray}
F_{T\rho,T}^{LL}&=&8\pi C_F m^4_B f_{D^*} r \int dx_B dx\int b_B db_B b db \phi_B \big\{\big[ \epsilon _T^{D^*}\cdot \epsilon _T^{\rho} [\sqrt{\zeta} (x (r^2-1) (\phi ^a-\phi ^v)+2 \phi ^v)\nonumber\\
&&+\zeta (2 x (r^2-1)+1)\phi ^T +(1-r^2) \phi ^T]-i \epsilon ^{n v \epsilon _T^{D^*}\epsilon _T^{\rho}} [\sqrt{\zeta} ((x (r^2-1)-2)\phi ^a\nonumber\\
&&-x (r^2-1)\phi ^v)+\zeta(2 x (r^2-1)+1)\phi ^T+(r^2-1)\phi ^T]\big]E_e(t_a)h_a(x_B,x,b,b_B)S_t(x)\nonumber\\
&&+\sqrt{\zeta}\big[\epsilon _T^{D^*}\cdot \epsilon _T^{\rho}[(\zeta-x_B-r^2+1)\phi ^v +(\bar{\zeta}+x_B-r^2)\phi ^a]+i \epsilon ^{n v \epsilon _T^{D^*}\epsilon _T^{\rho}} [ (\zeta-x_B -r^2\nonumber\\
&&+1)\phi ^a-(\zeta-x_B +r^2-1)\phi ^v]\big] E_e(t_b)h_b(x_B,x,b_B,b)S_t(|x_B-\zeta|)\big\},
\label{eq:f21}
\end{eqnarray}
\begin{eqnarray}
M_{T\rho,T}^{LL}&=&16\sqrt{\frac{2}{3}} \pi C_F m_B^4 \int dx_B dx dx_3 \int b_B db_B b_3 db_3 \phi_B \phi_{D^*}\big\{\big[ \epsilon _T^{D^*}\cdot \epsilon _T^{\rho}[\zeta ^{3/2} r_c (\phi ^a-\phi ^v) \nonumber\\
&& +\sqrt{\zeta} r_c((r^2-1) \phi ^a+(r^2+1)\phi ^v)+r (r^2-1) (x_B+x_3-1)\phi ^T-\zeta r ((r^2-1) \nonumber\\
&& \times (x_3+x)-2r^2+1)\phi ^T]-i \epsilon ^{n v \epsilon _T^{D^*}\epsilon _T^{\rho}} [\zeta ^{3/2} r_c (\phi ^a-\phi ^v)-\sqrt{\zeta} r_c ((r^2+1) \phi ^a\nonumber\\
&& +(r^2-1) \phi ^v)-r (r^2-1) (x_B+x_3-1)\phi ^T+\zeta r ((x_3-x)(r^2-1)+1)\phi ^T]\big] \nonumber\\
&& \times E_n(t_c)h_c(x_B,x,x_3,b_B,b_3)+r\big[\epsilon _T^{D^*}\cdot \epsilon _T^{\rho}[2 \sqrt{\zeta} (x_B+x (r^2-1)-x_3 \bar{\zeta}) \phi ^v\nonumber\\
&&+(r^2-1) (x_B-x\zeta-x_3 \bar{\zeta})\phi ^T]+i \epsilon ^{n v \epsilon _T^{D^*}\epsilon _T^{\rho}} [2 \sqrt{\zeta} (x_B+x (r^2-1)-x_3 \bar{\zeta}) \phi ^a \nonumber\\
&&+(r^2-1) (x_B+x\zeta -x_3 \bar{\zeta})\phi ^T ]\big]E_n(t_d)h_d(x_B,x,x_3,b_B,b_3)\big\},
\label{eq:m21}
\end{eqnarray}
\begin{eqnarray}
F_{TD^{*},T}^{LL}&=&8\pi C_F m^4_B f_\rho\sqrt{\zeta}\int dx_B dx_3\int b_B db_B b_3 db_3\phi_B\phi_{D^*} \big\{\big[ 
\epsilon _T^{D^*}\cdot \epsilon _T^{\rho}[x (r^2-1) (2\bar{\zeta}-r)\nonumber\\
&& +\bar{\zeta}+r^2+2r]-i \epsilon ^{n v \epsilon _T^{D^*}\epsilon _T^{\rho}} [x (r^2-1) (r-2\bar{\zeta})-\bar{\zeta}+r^2]\big]E_e(t_m)h_m(x_B,x_3,b_3,b_B) \nonumber\\
&& \times S_t(x_3)+r\big[\epsilon _T^{D^*}\cdot \epsilon _T^{\rho}[\zeta-x_B+2 r_c -r^2+1]-i \epsilon ^{n v \epsilon _T^{D^*}\epsilon _T^{\rho}} [\bar{\zeta}+x_B-r^2]\big] \nonumber\\
&& \times E_e(t_n)h_n(x_B,x_3,b_B,b_3)S_t(x_B)\big\},
\label{eq:f23}
\end{eqnarray}
\begin{eqnarray}
M_{TD^{*},T}^{LL}&=&16\sqrt{\frac{2}{3}} \pi C_F m_B^4\sqrt{\zeta}\int dx_B dx dx_3\int b_Bdb_Bbdb\phi_B\phi_{D^*} \big\{\big[ \epsilon _T^{D^*}\cdot \epsilon _T^{\rho}[r^2 (\bar{\zeta} ((2-x) \phi ^v\nonumber\\
&& -x \phi ^a)+(\bar{\zeta} x_3+x_B)(\phi ^a-\phi ^v))+ \bar{\zeta} x(\phi ^a+\phi ^v)]-i \epsilon ^{n v \epsilon _T^{D^*}\epsilon _T^{\rho}} [r^2 ((\bar{\zeta}(x-x_3)\nonumber\\
&& -x_B)\phi ^v +(\bar{\zeta}(x_3+x-2)+x_B)\phi ^a )-\bar{\zeta} x (\phi ^a+\phi ^v)] \big] E_n(t_o)h_o(x_B,x,x_3,b_B,b)\nonumber\\
&&+\big[ \epsilon _T^{D^*}\cdot \epsilon _T^{\rho}[ (r^2 (x_B-x_3 \bar{\zeta})-x \bar{\zeta}(r^2-1))\phi ^a+(x (r^2-1) (2 r-\bar{\zeta})-(r-2)\nonumber\\
&&\times r (x_B-x_3 \bar{\zeta}))\phi ^v ]+i \epsilon ^{n v \epsilon _T^{D^*}\epsilon _T^{\rho}} [(x (r^2-1) (2 r-\bar{\zeta})- r(r-2) (x_B-x_3 \bar{\zeta}))\phi ^a\nonumber\\
&& + (r^2 (x_B-x_3 \bar{\zeta})-\bar{\zeta}x (r^2-1))\phi ^v]\big] E_n(t_p)h_p(x_B,x,x_3,b_B,b)\big\},
\label{eq:m23}
\end{eqnarray}
\begin{eqnarray}
 F_{A\rho,T}^{LL,L}&=&8\pi C_F m^4_B f_B r\int dx_3dx\int bdbb_3db_3\phi_{D^*}\big\{\big[ 
\epsilon _T^{D^*}\cdot \epsilon _T^{\rho}[\sqrt{\zeta} (x (r^2-1) (\phi ^a-\phi ^v)-2 \phi ^v)\nonumber\\
&& -r_c(r^2-\zeta-1) \phi ^T]+i\epsilon ^{n v \epsilon _T^{D^*}\epsilon _T^{\rho}} [\sqrt{\zeta} (x(r^2-1)\phi ^v- (x(r^2-1)+2)\phi ^a) +(r^2-\bar{\zeta}) \nonumber\\
&& \times r_c \phi ^T] \big]E_a(t_e)h_e(x,x_3,b,b_3)S_t(x)+\sqrt{\zeta}\big[\epsilon _T^{D^*}\cdot \epsilon _T^{\rho} 
[(\bar{\zeta}x_3+\zeta-r^2 + 1)\phi ^v+(\bar{\zeta}x_3 \nonumber\\
&& +\zeta+r^2-1)\phi ^a]+i \epsilon ^{n v \epsilon _T^{D^*}\epsilon _T^{\rho}} [(\bar{\zeta}x_3+\zeta+r^2-1)\phi ^v+(\bar{\zeta}x_3+\zeta-r^2+1)\phi ^a]\big] \nonumber\\
&& \times E_a(t_f)h_f(x,x_3,b_3,b)S_t(|\bar{\zeta}x_3+\zeta|)\big\},
\label{eq:f22}
\end{eqnarray}
\begin{eqnarray}
M_{A\rho,T}^{LL}&=&16\sqrt{\frac{2}{3}} \pi C_F m_B^4\int dx_B dx dx_3\int b_Bdb_Bbdb\phi_B\phi_{D^*} \big\{\big[
\epsilon _T^{D^*}\cdot \epsilon _T^{\rho}[ (r^2 x_B(r^2-1)\nonumber\\
&& +\bar{\zeta}r^2((r^2-1)(x_3-1)+\zeta ) -\bar{\zeta}\zeta x (r^2-1) )\phi ^T-2 \sqrt{\zeta} r \phi ^v]+i \epsilon ^{n v \epsilon _T^{D^*}\epsilon _T^{\rho}}[(r^2 (\bar{\zeta}r^2\nonumber\\
&& -\bar{\zeta}^2-(\bar{\zeta}x_3+x_B)(r^2-1))-\bar{\zeta}\zeta x (r^2-1))\phi ^T-2 \sqrt{\zeta} r \phi ^a] \big] E_n(t_g)h_g(x_B,x,x_3,b,b_B)\nonumber\\
&& +(r^2-1) \big[\epsilon _T^{D^*}\cdot \epsilon _T^{\rho}[r^2 (x_B-x_3)+\zeta (r^2 (x_3-1)+x-1)+\zeta ^2 (1-x)]-i \epsilon ^{n v \epsilon _T^{D^*}\epsilon _T^{\rho}} \nonumber\\
&& \times [r^2 (x_B-x_3)+\zeta (r^2 (x_3-1)-\bar{\zeta}(x-1))]\big]\phi ^T E_n(t_h)h_h(x_B,x,x_3,b,b_B)\big\}.
\label{eq:m22}
\end{eqnarray}

The involved evolution factors $E_e(t)$, $E_a(t)$ and $E_n(t)$ are given by
\begin{eqnarray}
 E_e(t)&=&\alpha_s(t) \exp[-S_B(t)-S_{\rho}(t)],\nonumber\\
 E_a(t)&=&\alpha_s(t) \exp[-S_{D^{(*)}}(t)-S_{\rho}(t)],\nonumber\\
 E_n(t)&=&\alpha_s(t) \exp[-S_B(t)-S_{\rho}(t)-S_{D^{(*)}}(t)],
\end{eqnarray}
in which the Sudakov exponents are defined as
\begin{eqnarray}
S_B     &=& S(x_B\frac{m_B}{\sqrt2},b_B )+\frac53\int^t_{1/b_B}\frac{d\bar\mu}{\bar\mu} \gamma_q(\alpha_s(\bar\mu)),\\
S_{\rho}&=& S(x(1-r^2)\frac{m_B}{\sqrt2},b )+ S((1-x)(1-r^2)\frac{m_B}{\sqrt2},b )+ 2\int^t_{1/b}\frac{d\bar\mu}{\bar\mu} \gamma_q(\alpha_s(\bar\mu)),     \quad   \\ 
S_{D^{(*)}}&=& S(x_3(1-\zeta)\frac{m_B}{\sqrt2},b_3 ) +2\int^t_{1/b_3}\frac{d\bar\mu}{\bar\mu} \gamma_q(\alpha_s(\bar\mu)),
\end{eqnarray}
with the quark anomalous dimension $\gamma_q=-\alpha_s/\pi$. The explicit form for the function $s(Q,b)$ 
is~\cite{prd76.074018}
\begin{eqnarray}
s(Q,b)&=&\frac{A^{(1)}}{2\beta_{1}}\hat{q}\ln\left(\frac{\hat{q}}{\hat{b}}\right)-\frac{A^{(1)}}{2\beta_{1}}
               \left(\hat{q}-\hat{b}\right) \nonumber \\
&+&\frac{A^{(2)}}{4\beta_{1}^{2}}\left(\frac{\hat{q}}{\hat{b}}-1\right)-\left[\frac{A^{(2)}}
             {4\beta_{1}^{2}}-\frac{A^{(1)}}{4\beta_{1}}\ln\left(\frac
           {e^{2\gamma_E-1}}{2}\right)\right]\ln\left(\frac{\hat{q}}{\hat{b}}\right)\nonumber \\
&+&\frac{A^{(1)}\beta_{2}}{4\beta_{1}^{3}}\hat{q}\left[\frac{\ln(2\hat{q})+1}{\hat{q}}
       -\frac{\ln(2\hat{b})+1}{\hat{b}}\right]+\frac{A^{(1)}\beta_{2}}{8\beta_{1}^{3}}\left[\ln^{2}(2\hat{q})-\ln^{2}(2\hat{b})\right],
\end{eqnarray} 
with the variables
\begin{eqnarray}
     \hat q \ \ \equiv \ \mbox{ln}[Q/(\sqrt 2\Lambda)],\qquad \qquad  \hat b \ \ \equiv \ \mbox{ln}[1/(b\Lambda)], 
\end{eqnarray} 
and the coefficients $A^{(i)}$ and $\beta_i$ are
\begin{eqnarray}
 \beta_1&=&\frac{33-2n_f}{12},\qquad   \quad  \beta_2 \ = \ \frac{153-19n_f}{24},\nonumber\\
  A^{(1)}&=&\frac{4}{3},\qquad \qquad  \qquad A^{(2)}= \ \frac{67}{9}-\frac{\pi^2}{3}-\frac{10}{27}n_f+\frac{8}{3}    
                    \beta_1\mbox{ln}(\frac{1}{2}e^{\gamma_E}),
\end{eqnarray}
where $n_f$ is the number of the quark flavors and $\gamma_E$ is the Euler constant. 

The hard scale, denoted as $t_i$, are determined by selecting the maximum value of the virtuality associated with the internal momentum transition in the hard amplitudes, the specific expressions for the hard scales are given by : 
\begin{eqnarray}
  t_a&=&\max \{m_B\sqrt{|a_1|},m_B\sqrt{|a_2|}, 1/b, 1/b_B \},\nonumber\\
  t_b&=&\max \{m_B\sqrt{|b_1|},m_B\sqrt{|b_2|}, 1/b_B, 1/b \},\nonumber\\
  t_c&=&\max \{m_B\sqrt{|c_1|},m_B\sqrt{|c_2|}, 1/b_B, 1/b_3 \},\nonumber\\
  t_d&=&\max \{m_B\sqrt{|d_1|},m_B\sqrt{|d_2|}, 1/b_B, 1/b_3 \},\nonumber\\
  t_e&=&\max \{m_B\sqrt{|e_1|},m_B\sqrt{|e_2|}, 1/b, 1/b_3 \},\nonumber\\
  t_f&=&\max \{m_B\sqrt{|f_1|},m_B\sqrt{|f_2|}, 1/b_3, 1/b \};\nonumber\\
  t_g&=&\max \{m_B\sqrt{|g_1|},m_B\sqrt{|g_2|}, 1/b, 1/b_B \},\nonumber\\
  t_h&=&\max \{m_B\sqrt{|h_1|},m_B\sqrt{|h_2|}, 1/b, 1/b_B \};\nonumber\\
  t_m&=&\max \{m_B\sqrt{|m_1|},m_B\sqrt{|m_2|}, 1/b_3, 1/b_B \},\nonumber\\
  t_n&=&\max \{m_B\sqrt{|n_1|},m_B\sqrt{|n_2|}, 1/b_3, 1/b_B \},\nonumber\\
  t_o&=&\max \{m_B\sqrt{|o_1|},m_B\sqrt{|o_2|}, 1/b_B, 1/b \},\nonumber\\
  t_p&=&\max \{m_B\sqrt{|p_1|},m_B\sqrt{|p_2|}, 1/b_B, 1/b \}.
\end{eqnarray}
with the factors
\begin{eqnarray}
 a_1&=& (1-r^2 ) x, \qquad \qquad \quad \quad \quad a_2= (1-r^2 ) x_B x,\nonumber\\
 b_1&=& (1-r^2 ) (x_B-\zeta ), \qquad \qquad    b_2=a_2,\nonumber\\
 c_1&=&a_2, \qquad \qquad \qquad \qquad \qquad c_2= r_c^2-[(1-x)r^2+x][(1-\zeta)(1-x_3)-x_B],\nonumber\\
 d_1&=&a_2, \qquad \qquad \qquad \qquad \qquad d_2= (1-r^2)x [x_B-(1-\zeta ) x_3],\nonumber\\
 e_1&=& r_c^2-[1-x(1-r^2) ], \quad \quad \ \ \ e_2=(1-r^2)(1-x)[(\zeta-1)x_3-\zeta],\nonumber\\
 f_1&=& (1-r^2)[(\zeta-1)x_3-\zeta],\quad\;        f_2=e_2,\nonumber\\
 g_1&=&e_2, \qquad \qquad \qquad \qquad \qquad g_2= 1-[(1-x)r^2+x][(1-\zeta)(1-x_3)-x_B],\nonumber\\
 h_1&=&e_2, \qquad \qquad \qquad \qquad \qquad h_2= (1-r^2)(1-x)[(\zeta-1)x_3-\zeta+x_B],\nonumber\\
 m_1&=&(1-\zeta) x_3, \qquad \qquad \quad \quad \quad m_2=(1-\zeta) x_3 x_B;\nonumber\\
 n_1&=&  r_c^2-(r^2-x_B)(1-\zeta), \quad \quad n_2=m_2,\nonumber\\
 o_1&=&m_2, \qquad \qquad \qquad \qquad \qquad o_2=[(\zeta -1) x_3-\zeta ] [(1-x)(1-r^2)-x_B],\nonumber\\
 p_1&=&m_2, \qquad \qquad \qquad \qquad \qquad p_2=(1-\zeta) x_3  [x_B-(1-r^2)x ].
\end{eqnarray}

The threshold resummation factor $S_t(x)$ is of the form ~\cite{prd65.014007}:
\begin{eqnarray}
\label{eq-def-stx}
     S_t(x)=\frac{2^{1+2c}\Gamma(3/2+c)}{\sqrt{\pi}\Gamma(1+c)}[x(1-x)]^c,
\end{eqnarray}
with the parameter $c$ adopted to be $0.3$. 

The expressions of the hard functions $h_i$ with $i\in \{a,b,c,d,e,f,g,h,m,n,o,p\}$ are obtained through the Fourier 
transform of the hard kernel: 
\begin{eqnarray}
  h_i(x_1,x_2,x_3,b_1,b_2)&=&h_1(\beta,b_2)\times h_2(\alpha,b_1,b_2),\nonumber\\
  h_1(\beta,b_2)&=&\left\{\begin{array}{ll}
  K_0(\sqrt{\beta}b_2), & \quad  \quad \beta >0\\
  K_0(i\sqrt{-\beta}b_2),& \quad  \quad \beta<0
  \end{array} \right.\nonumber\\
  h_2(\alpha,b_1,b_2)&=&\left\{\begin{array}{ll}
  \theta(b_2-b_1)I_0(\sqrt{\alpha}b_1)K_0(\sqrt{\alpha}b_2)+(b_1\leftrightarrow b_2), & \quad   \alpha >0\\
  \theta(b_2-b_1)I_0(\sqrt{-\alpha}b_1)K_0(i\sqrt{-\alpha}b_2)+(b_1\leftrightarrow b_2),& \quad   \alpha<0 \quad
  \end{array} \right.
\end{eqnarray}
where $K_0$, $I_0$ are modified Bessel function with
$K_0(ix)=\frac{\pi}{2}(-N_0(x)+i J_0(x))$ and $J_0$ is the Bessel function,
$\alpha$ and $\beta$ are the factors $i_1, i_2$. 


\end{document}